\documentclass[aps,prb,twocolumn]{revtex4-2}
\usepackage{amssymb}
\usepackage{graphicx,amsmath}
\usepackage{hyperref}
\usepackage{color}
\usepackage{stix}
\newcommand{\bea}{\begin{eqnarray}}
\newcommand{\eea}{\end{eqnarray}}
\newcommand{\beq}{\begin{equation}}
\newcommand{\eeq}{\end{equation}}
\newcommand{\benu}{\begin{enumerate}}
\newcommand{\enu}{\end{enumerate}}

\newcommand{\ra}{\rangle}

\newcommand{\ga}{\gamma}

\newcommand{\ham}{\mathcal{H}}
\newcommand{\ord}{\mathcal{O}}

\newcommand{\cda}{c^{\dagger}}

\begin{document}

\title{
Hilbert space fragmentation imposed real spectrum of non-Hermitian systems}
\date{\today}
\author{Somsubhra Ghosh$^1$, K. Sengupta$^1$, and
Indranil Paul$^2$}
\affiliation{$^1$School of Physical Sciences, Indian Association for the
Cultivation of Science, Kolkata 700032, India. \\
$^2$Universit\'{e} Paris Cit\'{e}, CNRS, Laboratoire Mat\'{e}riaux
et Ph\'{e}nom\`{e}nes Quantiques, 75205 Paris, France.}

\begin{abstract}
We show that constraints imposed by strong Hilbert space fragmentation (HSF) along
with the presence of certain global symmetries can ensure
the reality of eigenspectra of non-Hermitian quantum systems; such a reality cannot
be guaranteed by global symmetries alone. We demonstrate
this insight for two interacting finite chains, namely the fermionic Nelson-Hatano and the
Su-Schrieffer-Heeger models, none of which has a $\mathcal{PT}$ symmetry. We show analytically that strong HSF and real spectrum are both
consequences of the same dynamical constraints in the limit of large interaction, provided the systems
have sufficient global symmetries. We also show that a local equal-time correlation function can detect
the many-body exceptional point at a finite critical interaction strength above which the
eigenspectrum is real.
\end{abstract}

\maketitle

%%%%%%%%%%%%%%%%%%%%%%%%%%%%%%%%%%%%%%%%%%%%%%%%%%%%%%%%%%%%%%%%
\section{Introduction}
\label{intro}

Non-Hermitian many-body Hamiltonians are of
great current interest for their relevance to open quantum systems,
and also for their novel properties without Hermitian
analog~\cite{reviews_nonH,
nonhlit1,nonhlit2,nonhlit3,nonhlit4,nonhlit5,nonhlit6,nonhlit7,nonhlit8,nonhlit9,
nonhlit10,nonhlit11,nonhlit12,nonhlit13,nhdyn1,nhdyn2,nhdyn3,nhdyn4,nhdyn5,nhdyn6}.
These features include novel topological properties including
the presence of so called exceptional
points. At these points, certain energy eigenvalues are degenerate
and the corresponding eigenfunctions
coalesce. Across such points, these eigenvalues
can transform from being real to complex~\cite{reviews_EP,
eptop1,eptop2,eptop3,eptop4,eptop5,eptop6}. Moreover, the non-equilibrium dynamics of such systems
are of interest since they posses qualitatively different characteristics compared to that
of their closed Hermitian counterparts \cite{nhdyn1,nhdyn2,nhdyn3,nhdyn4,nhdyn5,nhdyn6}. Some such
characteristics include the presence of athermal steady states and entanglement transitions following a quench
or in the presence of a periodic drive.

The purpose of the current work is to investigate an important related feature of these systems. Namely,
why the spectra of certain non-Hermitian Hamiltonians are entirely real in some parameter regimes.
Note, this question cannot be addressed completely by invoking global symmetries.
For example, pseudo-Hermitcity only guarantees that complex eigenvalues, if they appear,
come in complex conjugate pairs~\cite{Mostafazadeh2002a,Mostafazadeh2002b}.
Likewise, a so-called $\mathcal{P} \mathcal{T}$-symmetric system, where
 $\mathcal{P}$ and $\mathcal{T}$ refer to parity and time reversal operators, respectively,
has completely real eigenvalues
only in the regime where all the energy eigenfunctions are also simultaneously eigenfunctions
of the $\mathcal{P} \mathcal{T}$ operator~\cite{bender2007,zyablovsky2014,ozdemir2019},
and the question remains as to what guarantees the latter.
In fact, as discussed later, the models that we study in this work are not
$\mathcal{P} \mathcal{T}$-symmetric and, as such, $\mathcal{P} \mathcal{T}$ symmetry plays
no role.

In this work we show that the combination of dynamical constraints imposed in the limit of large
interaction along with global symmetries together can protect a ``phase'' where the spectrum
is entirely real. We illustrate our idea with finite chains of the fermionic Hatano-Nelson (HN)
and the non-Hermitian Su-Schrieffer-Heeger (SSH) models with nearest neighbor interaction.

An important ingredient in what follows is strong Hilbert space
fragmentation in the limit of infinitely large interaction, where the Fock space breaks
up into dynamically disjoint pieces whose number scales
exponentially with the system size~\cite{khemani2020,sala2020}. This
phenomena is the focus of intense research at present in Hermitian systems, since it leads
to non-ergodicity and the ability to generate exotic non-equilibrium
states~\cite{rakovszky2020,yang2020,tomasi2019,frey2022,hsf6,hsf7,hsf8,hsf9,hsf10,ghosh2023}.
%====================
\begin{figure}[t]
\begin{center}
\includegraphics[width=0.49\textwidth]{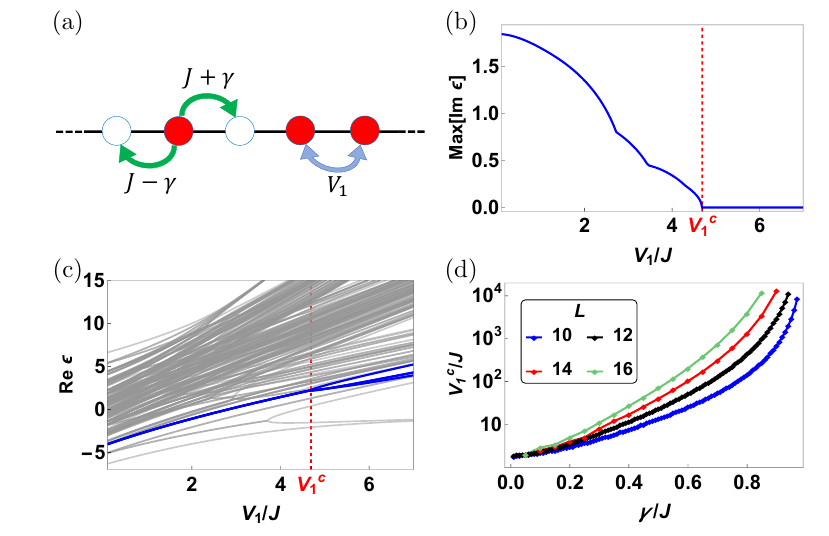}
\caption{(Color Online)  (a) Schematic of Hatano-Nelson model
with non-reciprocal hopping, and nearest neighbor interaction $V_1$.
Filled sites are marked red. (b), (c) show the maximum imaginary
and the real parts of the spectrum, respectively, as a function
of $V_1$, for system size $L=10$ and $\ga =0.2J$. At $V_1^c$
the system encounters a many-body
exceptional point. For $V_1 < V_1^c$ complex conjugate pairs of
eigenvalues first appear. In (c) the two eigenvalues which coalesce
at the exceptional point are delineated in blue. (d) $V_1^c$
diverges as $\ga \rightarrow J$, and also with $L$.}
\label{fig1}
\end{center}
\end{figure}
%====================

Our main results are the following. (1) We show analytically that  in the
fragmented limit there are many-body similarity transformations that map the above
non-Hermitian systems to Hermitian ones. Thus, in this limit,
the spectra are real. In fact, fragmentation and real spectrum are both shown
to be consequences of the
same dynamical constraints which emerge in the limit of infinitely large interaction.
(2) We show that the spectra persist to be real for arbitrarily large but finite interaction,
provided the systems have sufficient global symmetry protection. These
symmetries impose a hidden Hermiticity in those subspaces where the
reality of the spectrum is not
guaranteed by the fragmentation limit. Results (1) and (2) together
imply that the spectra are real for interactions above
finite critical values where the systems encounter many-body exceptional points.
(3) We compute a local equal-time density-density fermionic correlation
function and show that it can be used to detect the exceptional point location. Overall, our work
provides the first analysis of the role of dynamical constraints in
determining the spectrum of a non-Hermitian system.

The plan of the rest of the paper is as follows. In Sec.\ \ref{hnmodel}, we discuss the role
of HSF on reality of the eigenspectra of the Hatano-Nelson model and demonstrate the presence of a "phase" with entirely 
real eigenvalues. This is followed by Sec.\ \ref{SSH} where analogous phenomena is discussed for the 
SSH model. Next, in Sec.\ \ref{corr}, we discuss the behavior of the equal-time correlation function of the 
Hatano-Nelson model. Finally, we conclude in Sec.\ \ref{diss}. Some details of the calculations are 
elaborated in the appendices.

\section{Hatano-Nelson model}
\label{hnmodel}
\subsection{Model}

For pedagogical reason we demonstrate our principle in detail first for the HN model, and argue
later that the same holds for the SSH model. In terms of fermionic creation and annihilation
operators $(\cda_i, c_i)$ at site $i$ the HN Hamiltonian is
\beq
\label{eq:Hamiltonian} \ham_{HN} = \sum_{i=1}^{L} \left[ (J - \ga) \cda_i
c_{i+1} + (J + \ga) \cda_{i+1} c_i + V_1 \hat n_i \hat n_{i+1}
\right],
\eeq
where $\hat n_i \equiv \cda_i c_i$ is the number operator at site $i$,
$L$ is the system size, and $\ga >0$ is the  non-reciprocal hopping parameter, see
Fig.~\ref{fig1}(a). We study the system at
half-filling with $\sum_i n_i = L/2$, and we impose (anti) periodic
boundary condition for total particle number (even) odd. This choice
ensures that the system is traslationally invariant, as discussed below and in Appendix D. For open boundary
condition and $\ga < J$ the problem is trivial because there is a
one-body similarity transformation which makes $\ham_{HN}$
Hermitian~\cite{eptop3}. As discussed in detail in the next section,
$\ham_{HN}$ is pseudo-Hermitian, while its global symmetries are
$\mathcal{G} = (\mathcal{P} \mathcal{C}, \mathcal{R})$
with $[ \ham_{HN}, \mathcal{G}] =0$, where
$(\mathcal{P}, \mathcal{C}, \mathcal{R})$ are parity, charge
conjugation and translation by one site, respectively. Furthermore,
since integrability plays no role, our results are valid even in the
presence of next-nearest neighbor interaction.

The spectral properties of $\ham_{HN}$ are summarized in Fig.~\ref{fig1}, (b) - (d).
Panels (b) and (c) show that, for $\ga <J$, the spectrum is real for  $V_1 > V_1^c$, a
critical value. As shown in (c), at $V_1^c$ one pair (or two pairs) of eigenvalues
and eigenvectors coalesce at a many-body exceptional point, and
they become complex conjugate pairs for $V_1 < V_1^c$, as dictated by
the pseudo-Hermiticity of $\ham_{HN}$.
Panel (d) shows that $V_1^c$ diverges as $\ga \rightarrow J$, and
also with system size $L$.  Note, while these features
were reported recently~\cite{zhang2022}, the link between
Hilbert space fragmentation with the reality of the spectrum, which is
the focus of this work, has not been explored earlier.

\subsection{Global Symmetries}

We mention
two transformations which establish that the above Hamiltonian is
pseudo-Hermitian. First, parity transformation $\mathcal{P}$, such that the site number
$i \rightarrow (L-i +1)$. Since, this maps right hops to left hops and vice versa, we get
$\mathcal{P} \ham_{HN} \mathcal{P} = \ham_{HN}^{\dagger}$. Second, particle-hole transformation $\mathcal{C}$,
such that $\mathcal{C} c_i \mathcal{C} =(-1)^i\cda_i$. Under this transformation, $c_i^\dagger c_{i+1}\to -c_i c_{i+1}^\dagger = c_{i+1}^\dagger c_i$ and $c_{i+1}^\dagger c_i \to c_i^\dagger c_{i+1}$. Thus, the strengths of the right and the left hops are interchanged. The density-density interaction $n_i n_{i+1} \to (1-n_i)(1-n_{i+1})$. However, due to half-filling, this term remains invariant when summed over all sites. Thus, at half-filling we get
$\mathcal{C} \ham_{HN} \mathcal{C} = \ham_{HN}^{\dagger}$. However, by themselves these relations do
not guarantee that at large interaction $V_1$, the energy eigenvalues are real. It is well-known that
pseudo-Hermiticity simply guarantees that complex eigenvalues, if they appear, come in complex conjugate pairs.

Combining these two relations one gets a symmetry of the Hamiltonian $\ham_{HN}$, namely $\mathcal{P}\mathcal{C}$, since $(\mathcal{C}\mathcal{P})\ham_{HN}(\mathcal{P}\mathcal{C}) = \ham_{HN}$. It is due to this symmetry that some energy eigenvalues are doubly degenerate (the non-degenerate ones being eigenstates of $\mathcal{P}\mathcal{C}$) leading to situations when two pairs of eigenvalues coalesce at the exceptional point. In addition to this, $\ham_{HN}$ respects the translational symmetry $\mathcal{R}$ ($i \rightarrow (i+1) \text{ modulo } L$), which plays an important role in preserving the reality of the eigenspectrum for finite $V_1$, as explained in a later section.

Note, in the absence of spin degrees of freedom, the Hamiltonian of Eq.~\eqref{eq:Hamiltonian}
is trivially time reversal symmetric, with $\mathcal{T} \ham_{HN} \mathcal{T} = \ham_{HN}$.
This implies that $(\mathcal {P} \mathcal{T}) \ham_{HN} (\mathcal{P} \mathcal{T})
= \ham_{HN}^{\dagger}$. In other words, the model is not
$\mathcal{P} \mathcal{T}$-symmetric.

\subsection{Limit of Fragmentation}

For large interaction, we keep terms to linear order in $(J, \ga)$
and ignore those of order $(J, \ga)^2/V_1$ and smaller. This gives~\cite{dias2000},
\begin{align}
\label{eq:Hamiltonian2}
\ham_{HN} \approx \ham_{HN,f} &= \sum_{i=1}^{L} \left[\hat{P}_i \left(
(J - \ga) \cda_i c_{i+1}  \right. \right.
\nonumber \\
& \left. \left. + (J + \ga) \cda_{i+1} c_i \right) \hat{P}_i + V_1 \hat n_i \hat n_{i+1} \right],
\end{align}
where the projector $\hat{P}_i \equiv 1- (\hat n_{i-1} - \hat
n_{i+2})^2$ ensures that the hopping is constrained, and is allowed
only if the process does not change the total number of nearest
neighbor occupations $\hat{N}_d \equiv \sum_i \hat n_i \hat n_{i+1}$.

The Hermitian version of $\ham_{HN,f}$ has been shown to display strong Hilbert
space fragmentation~\cite{tomasi2019,frey2022}. Since fragmentation
is independent of whether the hopping mediated connectivity between
the many-body Fock states is reciprocal or not, the non-Hermitian
$\ham_{HN,f}$ shows fragmentation as well. Below we prove that the
dynamical constraints that give rise to fragmentation ensures the existence of a
a many-body similarity transformation that
maps $\ham_{HN,f}$ into a Hermitian form for $\ga < J$.

\subsection{Many-body Similarity Transformation}

The first step of the proof is
to label the many-body states. Traditionally, this is done using
``spins and movers''~\cite{dias2000,tomasi2019,frey2022}. Here we
take a different strategy, and we label them by ``defects''. A
``particle-defect'' and a ``hole-defect'' are two occupied or two
unoccupied nearest-neighbor sites, respectively.  Due to
half-filling, particle- and hole-defects appear in pairs, and their
numbers are conserved, since $\left[ \ham_{HN,f}, \hat{N}_d \right] =0$.
Thus, the Hilbert space factorizes into sectors with eigenvalue $N_d
= 0, 1, \ldots, L/2 -1$.

\begin{figure}[t]
\begin{center}
\includegraphics[width=0.49\textwidth]{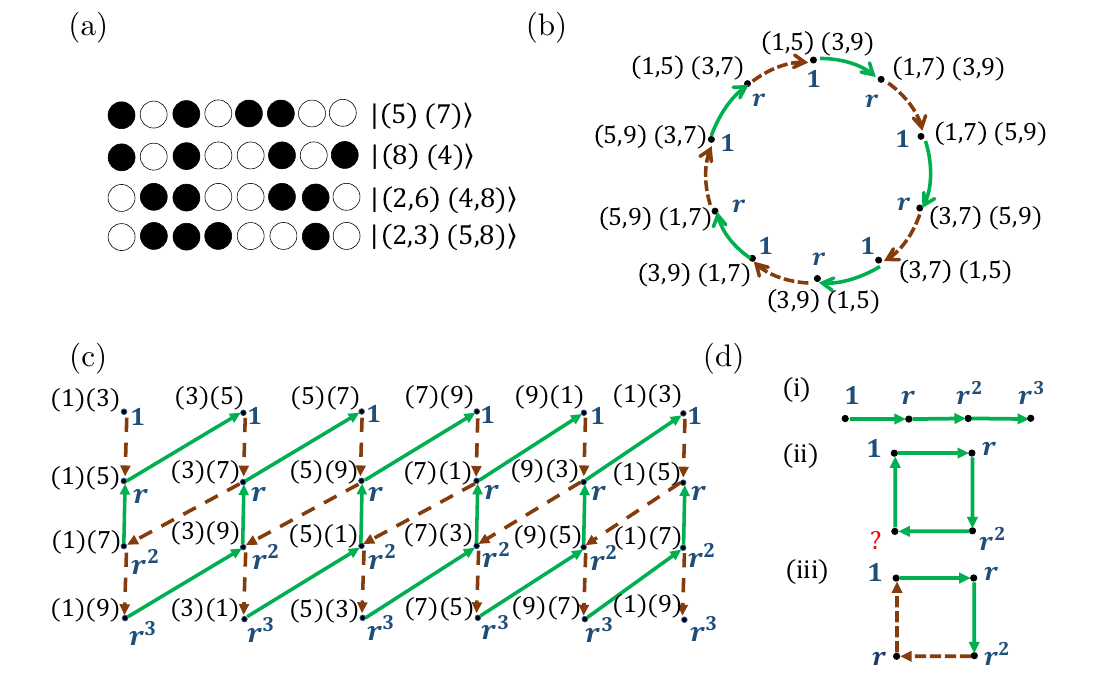}
\caption{(Color Online)  (a) Examples of labeling many-body
wavefunctions by the location of ``defects''. Two nearest neighbor
sites form a particle- or a hole-defect when they are both occupied
or both unoccupied, respectively. The defect position is the
location of the leftmost of the two sites. (b) and (c)
are examples of connectivities for $L=10$, $N_d = 2, 1$,
respectively, see text. Solid (green) and dashed (brown) arrows
denote fermions hopping to the right and left, respectively.
Reversing an arrow direction implies exchanging solid $\leftrightarrow$ dashed.
(d) Three examples of non-reciprocal
hopping over four sites. (i) has open boundary, while (ii) and (iii)
have periodic boundary conditions. $1, r, r^2$, \emph{et cetera} (in
blue) in (b)-(d) are the scaling factors $\lambda$, such as in
Eq.~\ref{eq:scaling}, of the wavefunctions next to them, which
define the similarity transformation, wherever possible. In (d, ii)
one of the sites, indicated by a question symbol, cannot be scaled
consistently. For a closed loop a similarity transformation is only
possible if there are equal number of solid and dashed arrows while
traversing the loop in a given direction, as in (d, iii).}
\label{fig2}
\end{center}
\end{figure}

All dynamically frozen (i.e., zero connectivity) states, which includes the
$N_d =0$ sector, have real
energies. For $N_d \neq 0$ we label a
defect position by the location of the leftmost of the two
nearest-neighbor sites. Thus, any state with $N_d=1$ has label $|(i)
(j) \ra$, where $i$ and $j$ are locations of the particle- and
hole-defect, respectively. Likewise, a state with $N_d =2$ is
labeled by $| (i_1, i_2) (j_1, j_2) \ra$, and $N_d = n$ by $| (i_1,
i_2, \ldots i_n) (j_1, j_2, \ldots j_n) \ra$. Since the fermions are
indistinguishable, permutations of $i$ and of $j$ imply the same
state. Thus, the state $|(5) (7) \ra$ shown in Fig.~\ref{fig2}(a), belongs to
$N_d =1$, and has a particle- and a hole-defect at sites $i = 5, 7$, respectively.

Due to half-filling the defect locations are not arbitrary, but follow certain rules. (a)
If two particle-defects at $i_1$ and $i_2$ are ``adjacent'', then
$(i_1, i_2)$ can only be (odd, even) or (even, odd). The same
applies for two adjacent hole-defects. Here, ``adjacent'' does not
imply defects located right next to one another. Two defects are
adjacent if there is no third defect in between the two while
traversing in one of the two directions.
(b) If a particle-defect at $i_1$ is adjacent to a hole-defect at $j_1$,
then $(i_1, j_1)$ can only be (even, even) or (odd, odd). One can
verify that the wavefunctions in Fig.~\ref{fig2}(a) satisfy these
rules.

%====================
%====================

The second step is to determine the defect dynamics which, due to
the constrained hopping, obey the following rules. (i) An allowed
fermion hop changes $i$ or $j$ by $\pm$2 modulo $L$. (ii) Since
second nearest neighbor hopping is absent, two defects cannot cross each other.
It is shown in Appendix A that due to
rules (i) and (ii) each sector of $N_d$ breaks into
exponentially large number of disjoint subsectors, i.e. fragments,
that scale as $e^{N_d}$ \cite{sala2020}.

The third step is to establish the constrained hopping induced
connectivity between the many-body wavefunctions within each
non-trivial subsector. There is no general pattern for these
connectivities, and they need to be worked out case by case,
even though the proof below holds for all the connectivities.
To show few examples,  Fig.~\ref{fig2}(b)
is the connectivity for $L = 10$, $N_d = 2$ with $(i_1, i_2)(j_1,
j_2) =$ (odd, odd)(odd, odd), while Fig.~2(c) is for $L = 10$, $N_d =
1$ with $(i)(j) =$ (odd)(odd). The dashed and solid arrows denote
fermions hopping to the left and right (amplitudes $J_{1,2} \equiv J \mp \ga$),
respectively.
Reversing an arrow implies $J_1 \leftrightarrow J_2$. A
fermion hopping left can result either a particle-defect to move
left, i.e., $i \rightarrow (i -2)$ mod $L$, or a hole-defect to move
right, i.e., $j \rightarrow (j +2)$  mod $L$. Thus, each connectivity
diagram can be viewed as a single
``particle'' hopping in the abstract space of many-body
wavefunctions in a non-reciprocal manner.

The fourth and final step of the proof is to establish the existence
of the similarity transformation in each sub-sector. For pedagogical
reason we first consider few examples of non-reciprocal hopping of a
single particle in a four-site system. Fig.~\ref{fig2}(d, i) is a
linear chain with open boundary condition. This can be mapped to a
Hermitian form for $\ga < J$ by the scaling
\beq
\label{eq:scaling}
|i \rangle \rightarrow \lambda_i |i \rangle, \quad \langle i |
\rightarrow (1/\lambda_i) \langle i |,
\eeq
where $\lambda_i = 1, r,
r^2, r^3$, for $i = 1, \ldots, 4$, respectively, and $r \equiv
\sqrt{J_2/J_1}$~\cite{eptop3}. However, for periodic boundary
condition, as in Fig.~\ref{fig2}(d, ii), the transformation will not
work since the new link 4 -1 of the closed loop cannot be made Hermitian.
This exemplifies that finding similarity transformations is non-trivial
where the connections form closed loops, which is precisely our case as seen in, e.g.,
Fig.~\ref{fig2}(b, c). Now consider Fig.~\ref{fig2}(d, iii) which is
also a closed loop, but where the hops are $J_2, J_2, J_1, J_1$
moving clockwise. In this case, once again, a similarity mapping
exist, with $\lambda_i = 1, r, r^2, r$, respectively.  This example
illustrates the crucial property that a closed loop with \emph{equal} number
of $J_1$ and $J_2$ hops, while traversing along a direction,
can be mapped to a Hermitian form. This is because
a $J_2$ link requires an additional scaling of $r$ for the
second site compared to the first, which is compensated by a $J_1$ link which requires a
$1/r$ scaling. This is exactly the case for the connectivities of
Fig.~\ref{fig2} (b, c), where the scalings associated with the
wavefunctions are marked in blue. Additional examples of such
scalings are discussed in the Appendix B. We prove below that
\emph{all} the connections of $\ham_{HN, f}$ are such that each and every
possible loop has this property.

Starting from any $| (i_1, i_2, \ldots i_n) (j_1, j_2, \ldots j_n) \ra$,
a closed loop is obtained in three basic ways.

(1) If one or more of the site indices change as, say,  $i
\rightarrow i^{\prime} \rightarrow i^{\prime \prime}$ and so on, and
then reverse the path to go back to $i$, while obeying the rules (i)
and (ii). Since the reverse of a $J_1$ hop is a $J_2$ hop, and vice
versa, traversing the loop along one direction will necessarily have
equal number of $J_1$ and $J_2$ hops. The loop (1)(7) $\rightarrow$
(1)(5) $\rightarrow$ (3)(5) $\rightarrow$ (3)(7) $\rightarrow$
(1)(7) in Fig.~\ref{fig2}(c) is an example.

(2) If a defect does not retrace its path, but moves across the
chain, traversing $L$ sites, and gets back to its original position
using the periodic boundary condition. However, according to rule
(ii) this can happen only if all other defects perform the same
circular motion in the same direction and regain their original
positions, each having traversed $L$ sites. Since a particle-defect
moving to the right is a $J_2$ hop, while a hole-defect moving to
the right is a $J_1$ hop, and since there are equal number of
particle- and hole-defects, this loop, too, will have equal number
of $J_1$ and $J_2$ hops. Starting from state (1)(3) on the left side
of Fig.~\ref{fig2}(c) and ending again at (1)(3) on the right side
of the figure is an example of such a loop.

(3) In some cases, such as in Fig.~\ref{fig2}(b), it is possible for
the defects to exchange positions such that
$i_1 \rightarrow i_2 \rightarrow i_3 \ldots \rightarrow i_n \rightarrow i_1$, and
$j_1 \rightarrow j_2 \rightarrow j_3 \ldots \rightarrow j_n \rightarrow j_1$. In this case a loop is completed by permuting the indices,
while the defects neither retrace their paths nor complete the
circle. Here the sum of the sites traversed by all the
particle-defects is $L$ and the same is true for all the
hole-defects, and they are along the same direction. Thus, again
here the loop has equal number of $J_1$ and $J_2$ hops.

This completes the proof that $\ham_{HN,f}$ can be mapped into a
Hermitian form for $\ga < J$; this feature guarantees
reality of eigenspectrum of ${\mathcal H}_{HN,f}$ in this limit.

\subsection{Finite $V_1^c$ and Symmetry Protection}

The above conclusion is not sufficient for our purpose once $V_1$ is large but finite.
To understand why, consider two eigenstates of $\ham_{HN,f}$ from the
same sub-sector of $\hat{N}_d$. Measuring energies from the average eigenvalue, the sub-system
has the structure
\[
\ham_{eff} =
\begin{pmatrix}
l & m_1 + m_2 \\
m_1 - m_2 &  - l
\end{pmatrix},
\]
with eigenvalues $ \pm \sqrt{l^2 + m_1^2 - m_2^2}$. Since
$m_{1,2} \sim \ord((J, \ga)^2/V_1)$ or smaller, for finite $l$ the reality of the eigenvalues
is guaranteed for $V_1$ sufficiently large. But, this argument fails in case of degeneracy when
 $l=0$. Nevertheless, the reality of the spectrum can still be
protected if the two degenerate states are connected by a symmetry $\mathcal{G}$ of the full Hamiltonian $\ham_{HN}$, which squares to identity in the degenerate subspace, as we show below.%discussed in the main text and provide one concrete example to illustrate the point.

Let $|\psi\rangle$ and $|\phi\rangle$ represent two degenerate right eigenstates of $\ham_{HN,f}$ connected by a symmetry $\mathcal{G}$ of both $\ham_{HN,f}$ and $\ham_{HN}$, i.e. $|\phi\rangle= \mathcal{G}|\psi\rangle$. The corresponding right eigenbras are $\langle\psi|$ and $\langle\phi|=\langle\psi|\mathcal{G}^\dagger$. Further, let us assume that $\mathcal{G}^2 |\psi\rangle = |\psi\rangle$ and $\mathcal{G}^2 |\phi\rangle=|\phi\rangle$. The off-diagonal matrix elements are seen to be equal as follows.
\begin{eqnarray}
    \langle\psi|\ham_{HN}|\phi\rangle = \langle\phi|\mathcal{G}\ham_{HN}\mathcal{G}|\psi\rangle&=&\langle\phi|\ham_{HN}\mathcal{G}^2|\psi\rangle\nonumber\\&=&\langle\phi|\ham_{HN}|\psi\rangle
    \label{eq:symprot}
\end{eqnarray}
Since the eigenvectors of $\ham_{HN,f}$ can be chosen to be real when its eigenvalues are real, these matrix elements are also real. This implies that the full Hamiltonian $\ham_{HN}$ effectively has a Hermitian structure within this degenerate subspace. The degeneracy of the two states $|\psi\rangle$ and $|\phi\rangle$ at the level of $\ham_{HN,f}$ is lifted by $\delta \ham_{HN}=\ham_{HN}-\ham_{HN,f}$ in a hermitian manner at most. This leads to level repulsion which is opposite to what is required for two eigenstates to coalesce. This also implies that eigenstates, or their linear combinations, that are related by symmetry of the type of $\mathcal{G}$ cannot coalesce. Thus, there will be a finite range of $V_1$ for which the spectrum of $\ham_{HN}$ would be real, until one (or two) pair(s) of eigenstates, not related by any symmetry (of the type of $\mathcal{G}$), coalesce at an exceptional point. We illustrate this using one concrete example in Appendix C.

As discussed in Appendix D, the above symmetry protection can be destroyed
by a choice of boundary condition that breaks translation symmetry.
In this case one has complex eigenvalues
for any finite value of $V_1$, even though the spectrum is real in the fragmented limit.

We estimate $l \sim \sqrt{J^2- \ga^2}/e^{cL}$ by the average level spacing of a sub-sector,
where the constant $c > 0$ depends on the sub-sector size.
Empirically, we find that, for the pair that coalesce, $m_1$ is at least
one order of magnitude smaller than $m_2$, while
$m_2 \propto V_1^{-\alpha} (J^2 - \gamma^2)^{-\beta/2}$
where the exponents $(\alpha, \beta)$ are $L$-dependent. This implies that
$V_1^c \sim J e^{cL/\alpha}/(1- (\gamma/J)^2)^{(\beta + 1)/(2\alpha)}$.
Thus, as shown in Fig.~\ref{fig1}(d), $V_1^c$ diverges exponentially with $L$,
and as a power-law with a $L$-dependent exponent for
$\gamma \rightarrow J$.

Note, in passing, that for certain values of $L$ the two coalescing
levels at $V_1^c$ are each doubly degenerate, so that below $V_1^c$
there are two pairs of complex conjugate eigenvalues. This
degeneracy is related to $\mathcal{P} \mathcal{C}$ invariance.

\section{Generalization to Non-Hermitian Su-Schrieffer-Heeger model}
\label{SSH}

\subsection{Model}
\begin{figure}
\includegraphics[width=1.0\linewidth]{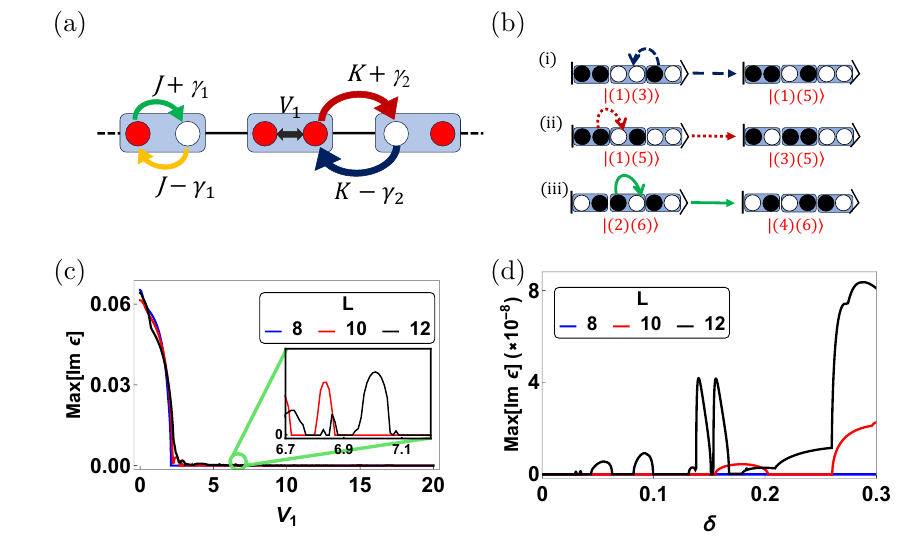}
\caption{(a) Schematic of the interacting SSH model with non-reciprocal intra- and inter-cellular hops and nearest neighbor interaction $V_1$. The blue boxes indicate one unit cell of the SSH chain and the filled sites are marked in red. (b) Defects at odd site can only move by an intercellular hop (blue dashed and red dotted in (i) and (ii)), while defects at even site can move by an intracellular hop (green solid in (iii)). (c) Variation of the maximum imaginary part of the eigenvalues with the interaction strength $V_1$ for three system sizes $L=8, 10, 12$ corresponding to $\ga_1=\ga_2=0.1$, $J=0.9$ and $K=1.0$. For all the system sizes, beyond a certain interaction strength, the imaginary parts of the eigenvalues become zero within numerical precision. The inset shows islands at $V_1<V_1^c$ (for instance, $V_1^c=23.8$ for $L=10$) for $L=10$ (red) and $L=12$ (black), where the imaginary part is finite, but orders of magnitudes smaller than the real part. (d) Variation of the maximum imaginary part of the eigenvalues with $\delta=K-J$ for three system sizes $L=8,10,12$ with $\ga_1=\ga_2=0.1$, $K=1.0$ and $V_1=1000$. The presence of a critical value of $\delta$ ($\delta_c =0.156$ for $L=10$, for example), below which the spectrum is completely real is seen. Also, the value of $\delta_c$ at a given $V_1$ reduces with increase in system size. See text for details.}
\label{fig:fig4}
\end{figure}

Next we discuss the non-Hermitian interacting SSH model as one possible generalization of the concept presented in the previous section. We consider a fermionic SSH chain with left (right) intracellular hopping $J_1=J-\gamma_1$ ($J_2=J+\gamma_1$), left (right) intercellular hopping $K_1=K-\gamma_2$ ($K_2=K+\gamma_2$) and a nearest neighbor interaction $V_1$, as shown in Fig. \ref{fig:fig4}(a). The Hamiltonian of the system reads
\begin{align}
\label{eq:nhssh}
\ham_{SSH}  = &\sum_{i=1}^{L/2} \left[ J_1 \cda_{2i-1} c_{2i}
+ J_2 \cda_{2i} c_{2i-1}  \right.
\nonumber\\
&+ \! \left.  K_1 \cda_{2i} c_{2i+1} \! + \! K_2 \cda_{2i+1} c_{2i} \right]
\! + \! \sum_{i=1}^{L} V_1 \hat{n}_i \hat{n}_{i+1}.
\end{align}
Each unit cell is composed of two non-equivalent sites $2i-1$ and $2i$ so that there are $L$ sites in total. The nearest neighbor interaction $V_1$ does not distinguish between the two non-equivalent sites. The non-Hermiticity of $\ham_{SSH}$ arises due to the non-reciprocal intracellular and intercellular hoppings.

As we show here, there exists a one-body similarity transformation by virtue of which for any parameter set
$(J,K,\ga_1,\ga_2)$, $\gamma_1$ can be made equal to $\gamma_2$, while keeping different values of $J$ and $K$. To do so, we scale the particle creation and annihilation operator on every odd site by
\[
c_{2j-1}^{\dagger} \to \lambda c_{2j-1}^{\dagger}, \quad c_{2j-1} \to \frac{1}{\lambda} c_{2j-1}
\]
respectively. In terms of this transformed operators, the Hamiltonian reads
\begin{align}
\label{eq:nhsshtransformed}
\ham_{SSH}  = &\sum_{i=1}^{L/2} \left[ \Tilde{J}_1 \cda_{2i-1} c_{2i}
+ \Tilde{J}_2 \cda_{2i} c_{2i-1}  \right.
\nonumber\\
&+ \! \left.  \Tilde{K}_1 \cda_{2i} c_{2i+1} \! + \! \Tilde{K}_2 \cda_{2i+1} c_{2i} \right]
\! + \! \sum_{i=1}^{L} V_1 \hat{n}_i \hat{n}_{i+1}.
\end{align}
where $\Tilde{J}_1=\Tilde{J}+\Tilde{\ga}_1=\lambda J_1$, $\Tilde{J}_2=\Tilde{J}-\Tilde{\ga}_1= J_2/\lambda$, $\Tilde{K}_1=\Tilde{K}+\Tilde{\ga}_2=K_1/\lambda$ and $\Tilde{K}_2=\Tilde{K}-\Tilde{\ga}_2=\lambda K_2$. Choosing
\[
\lambda =\sqrt{\frac{J+K+\ga_2-\ga_1}{J+K-\ga_2+\ga_1}},
\]
one gets $\Tilde{\ga}_1=\Tilde{\ga_2}$ in this transformed frame, while $\Tilde{J}\neq \Tilde{K}$. For our numerical results, we choose our parameter set to obey this and vary $\delta=K-J$, without any loss of generality.

 We consider the chain at half-filling, $\sum_i n_i=L/2$ with (anti-)periodic boundary condition for (even) odd $L/2$.  Fig. \ref{fig:fig4}(c) shows that in this case too, there is a strength of the nearest neighbor interaction, $V_1^c$ above which, the spectrum is completely real within numerical precision. For $V_1<V_1^c$, islands appear in the spectrum where the imaginary part is orders of magnitudes smaller than the real part of the eigenvalues. The islands progressively shorten in height before hitting zero within numerical precision at $V_1^c$. The inset of Fig. \ref{fig:fig4}(c) shows such islands. The value of this interaction strength is seen to increase with system size as well as with $\delta$.
%%%%%%%%%%%%%%%%%%%%%%%%%%%%%%%%%%%%%%%%%%%%%%%%%%%
%%%%%%%%%%%%%%%%%%%%%%%%%%%%%%%%%%%%%%%%%%%%%%%%%%%

\subsection{Fragmentation and Similarity Transformation}

We consider the large $V_1$ limit, $V_1\gg J, K, \gamma_1, \gamma_2$ and focus on the first-order effective Hamiltonian
 \begin{eqnarray}
    \ham_{SSH,f}&=&\sum_{i=1}^{L/2} \Big[ \hat{P}_{2i} \big(J_1 \cda_{2i-1} c_{2i}
+ J_2 \cda_{2i} c_{2i-1}\big)\hat{P}_{2i}\nonumber\\ &+& \hat{P}_{2i+1}\left(K_1 c_{2i}^\dagger c_{2i+1}+ K_2 c_{2i+1}^\dagger c_{2i}\right) \hat{P}_{2i+1}\nonumber\\
    &+& V_1 \sum_{i=1}^L \hat{n}_{i}\hat{n}_{i+1} \Big]
    \label{eq:nhsshfrag}
\end{eqnarray}
where $\hat{P}_i=1-\left( \hat{n}_{i-2}-\hat{n}_{i+1}\right)^2$ is a projector that only allows those hops which preserve the number of nearest neighbor pairs (defects), $N_d=\sum_i n_{i}n_{i+1}$. It is useful to remember here that a particle-defect refers to two particles sitting on neighboring sites and a hole-defect refers to two neighboring vacant sites. In addition, since a particle hop changes the position of a defect by $\pm 2$ modulo $L$ sites, therefore the number of defects at odd sites $\left( N_{d,odd}=\sum_i^{'} n_i n_{i+1}\right )$ and even sites $\left(N_{d,even}=\sum_i^{''} n_{i}n_{i+1}\right)$ are individually conserved by $\ham_{SSH,f}$. Here, single (double) prime implies sum over odd (even) sites only. It is easy to see now that these constraints are similar to those which appear in the Hatano-Nelson model in the large $V_1$ limit. Since the connectivity of states in the Hilbert space does not depend on the hopping strengths (as long as they are much smaller than $V_1$), therefore $\ham_{SSH,f}$ also exhibits strong Hilbert space fragmentation in spite of having different intercellular and intracellular hoppings.

In addition, it is important to note the following. (i) At half-filling, the number of particle defects at the odd sites is equal to the number of hole defects at odd sites and the same holds true for defects at even sites. (ii) A defect at an odd site can only move by a intercellular hop and a defect at an even site can only move by a intracellular hop (Fig. \ref{fig:fig4}(b)). (iii) Two defects cannot cross each other.

It remains to understand whether the similarity transformation goes through in the fragmented limit. For this purpose, we need to label the states first. The labelling scheme is similar to that used in the case of Hatano-Nelson model. %, with the only difference that now a defect position is specified by both the unit cell and the sublattice to which it belongs.%
Thus, in the $N_d=1$ sector, $|(i)(j)\rangle$ denotes a state having a particle-defect at site $i$ and a hole-defect at site $j$. Similarly, in the $N_d=2$ sector, a state is labelled as $|(i_1,i_2)(j_1,j_2)\rangle$ and a state in the $N_d=n$ sector is $|(i_1,i_2,\dots ,i_n)(j_1,j_2,\dots ,j_n) \rangle$.

Next, it is again useful to remember here that for the similarity transformation to be consistent along a closed loop (in Fock space), the number of right non-reciprocal hops should be equal to the number of left non-reciprocal hops along the loop. Since in this particular case, the intercellular and intracellular hops are individually non-reciprocal, this criterion translates to having equal number of right and left intercellular hops and equal number of right and left intracellular hops separately along any closed loop. To check whether this is the case, we again consider the three ways by which a state $|(i_1,i_2,\dots ,i_n)(j_1,j_2,\dots ,j_n) \rangle$ can close on to itself in Fock space.

(1) \textbf{The defects retrace their path}: In this case, one or more defects hop following $i\to i'\to i''\to\dots$ and then retrace their path to return to the original configuration and close the loop. Since a right intercellular (intracellular) hop will necessarily require a left intercellular (intracellular) hop for its reversal, therefore in this case, it trivially holds that a closed loop will have equal number of right and left intercellular (intracellular) hops.

\begin{figure}
\includegraphics[width=0.75\linewidth]{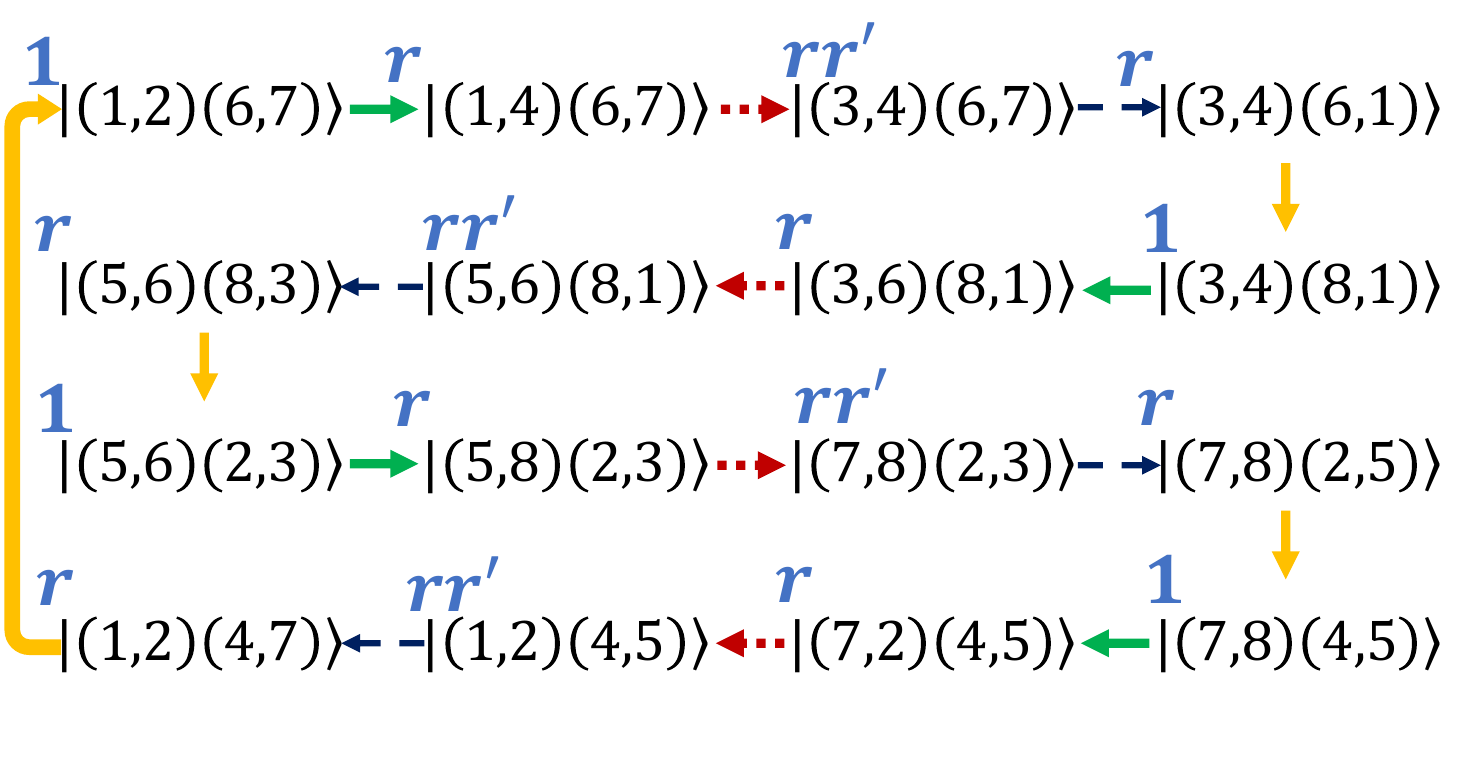}
\caption{Connectivity diagram of $\ham_{SSH,f}$ in Fock space showing type (2) connectivity in which the defects traverse through the entire chain to return to their original configuration. Solid green and yellow arrows refer to right and left intracellular hops respectively while red dotted and blue dashed arrows refer to right and left intercellular hops respectively. See text for details.}
\label{fig:fig5}
\end{figure}

(2) \textbf{The defect moves across the chain making $L$ hops and returns to its original position using the periodic boundary condition}: Due to rule (iii) above, this implies that all other defects would also have to return to their initial positions after going across the chain in the same direction. Each such defect will have to make $L$ hops across the chain. From rule (i) and (ii), this implies that there will be equal number of particle- and hole-defects undergoing intercellular (intracellular) hops in the same direction. Since a particle-defect hopping to the right is equivalent to a particle hopping to the right and a hole-defect hopping to the right is equivalent to a particle hopping to the left, this suggests that in this case too, there will be equal number of right and left intercellular hops and equal number of right and left intracellular hops separately along a closed loop. This is shown in Fig. \ref{fig:fig5}.

(3) \textbf{The defects permute their positions to reach the initial configuration}: In this case, a sequence of defects exchange their positions as $i_1 \to i_2 \to i_3 \to \dots \to i_n \to i_1$ and $j_1 \to j_2 \to j_3 \to \dots \to j_n \to j_1$ to reach the initial state and complete the loop. Thus, the sum of the sites traversed by all the particle-defects is
$L$, and the same is true for all the hole-defects.
However this type of connectivity is only possible if all the defects are present either
on the odd sites or on the even sites. If the defects are on odd (even) sites and the
motion is
along the anticlockwise direction, then all the hops of the particle-defects will be of the
type $K_2$ ($J_2$), while those of the hole-defects will be of the type $K_1$ ($J_1$).
Thus, here too there will be an equal number of $K_{1,2}$ hops and no $J_{1,2}$ hop,
or vice versa.

Thus, a consistent similarity transformation can always be carried out along any closed loop with each right and left intracellular (intercellular)
hop being scaled by a factor $r$ ($r'$) and $1/r$ ($1/r'$) respectively, where $r=\sqrt{\frac{J_2}{J_1}}$ and $r'=\sqrt{\frac{K_2}{K_1}}$.

%%%%%%%%%%%%%%%%%%%%%%%%%%%%%%%%%%%%%%%%%%%%%%%%%%%
%%%%%%%%%%%%%%%%%%%%%%%%%%%%%%%%%%%%%%%%%%%%%%%%%%%

\subsection{Symmetry Protection}

In this section, we discuss the protection of the real eigenvalues once we move out of the limit
of fragmentation $\mathcal{H}_{SSH,f}$ by considering large, but finite, $V_1$. As discussed for the HN model, the nondegenerate eigenvalues are guaranteed to stay
real once $V_1$ is finite, while the degenerate ones are not. In the HN model we found that the degenerate subspaces are symmetry protected which enforces Hermiticity in
these subspaces. The case of the SSH model is more complex because, due to
period doubling with $J \neq K$, translations by odd number of sites are no longer
symmetry operations. Thus,
we find certain pairs of degenerate eigenvalues in the fragmented limit which are not symmetry
connected. These pairs are symmetry connected only if $J = K$ when the
full translation symmetry of the chain is restored, i.e., in the limit of the HN model.
As expected, we find that in such projected non-symmetry connected two-dimensional subspaces the effective Hamiltonian is non-Hermitian with the structure
\[
\ham_{eff}=
\begin{pmatrix}
l & m_1 + m_2 \\
m_1 - m_2 &  - l
\end{pmatrix},
\]
with $m_{1,2} \neq 0$. However, if we define $\delta \equiv (K - J)$ as the deviation from
the HN limit, we are guaranteed that the non-Hermitian component $m_2 \rightarrow 0$
as $\delta \rightarrow 0$. Thus,  for $V_1$ sufficiently large (larger than $V_{1}^{c}$ of the
HN model) there is a finite critical value $\delta_c$ below which $m_1 > m_2$
for all such non-symmetry connected degenerate
pairs of eigenvalues, which guarantees that the eigenvalues stay real
even if the subspace
is non-Hermitian. This is shown in Fig. \ref{fig:fig4}(d) for $L=8,10,12$. The value of $\delta_c$ is seen to decrease with increase in system size for a given value of $V_1$. Beyond $\delta_c$, island-like features again emerge suggesting that the behavior of $m_2/m_1$ is non-monotonic as a function of $\delta$.
%====================
\begin{figure}[t]
\begin{center}
\includegraphics[width=0.49\textwidth]{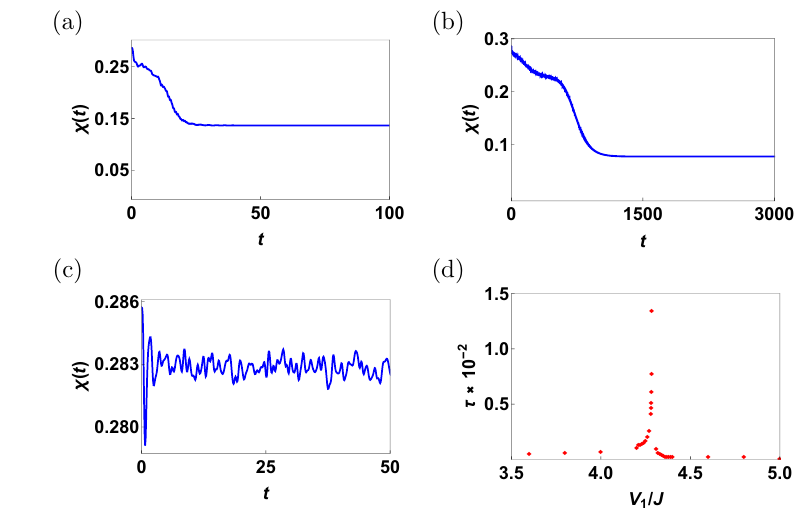}
\caption{(Color Online)  (a)-(c) Time evolution of the correlation function $\chi(t)$ for
$V_1$ = 3J, 4.285J, and 6J, respectively, for $L=14$ and $\gamma =0.2J$. In (b) the system is
very close to the exceptional point at $V_1^c \approx 4.2863$. (d) Variation of the relaxation
timescale $\tau$ with $V_1$ showing a one-sided divergence as $V_1 \rightarrow V_1^c$
from below.
}
\label{fig3}
\end{center}
\end{figure}
%====================

\section{Correlation function and detection of Exceptional Point}
\label{corr} 

In this section, we study the time evolution
of a correlation function which, in principle, can be measured to
identify the location of a many-body exceptional point. We choose the Hatano Nelson model for this purpose. One such
example is
$\chi(t) =\langle \psi(t)|\hat N_d|\psi(t)\rangle/L$
starting from a random Fock state $|\psi(0)\rangle =\sum_m c_m |\phi_m\rangle$,
expanded in the basis of the right eigenvectors $|\phi_m\rangle$ of $\ham_{HN}$.
The time-evolved wavefunction, suitably normalized to account for the non-Hermiticity of
$\ham_{HN}$, is
\[
|\psi(t)\rangle = \frac{e^{-i \ham _{HN}t/\hbar} |\psi(0)\rangle}{||e^{-i
\ham_{HN} t/\hbar} |\psi(0)\rangle||} = \frac{\sum_m
c_m(t)|\phi_m\rangle}
{\sqrt{\sum_{m,n}c_m^*(t)c_n(t)\langle\phi_m|\phi_n\rangle}},
\]
where $c_m(t)=c_m e^{-i\epsilon_m t/\hbar}$.

Fig.~\ref{fig3}, panels (a)-(c) show the time evolution of $\chi(t)$
for $V_1$ less than, nearly equal to, and greater than $V_1^c$,
respectively, for the HN model. For $V_1 < V_1^c$ the time evolution is dominated by
the eigenvalue with the largest imaginary component. Consequently,
after a timescale $\tau \sim 1/{\rm Max[Im} \, \epsilon]$, the
correlation function attains a steady state value $\chi(t \gg \tau) \sim 1/L \langle\phi_m^*|\hat
N_d|\phi_m^*\rangle$, where $|\phi_m^*\rangle$ is the eigenvector
with the largest ${\rm Im} \, \epsilon$. This implies that $\tau$
diverges as $V_1 \rightarrow V_1^c$ from below, as seen in
Fig.~\ref{fig3}(d). For $V_1 \geq V_1^c$ all the eigenvalues are
real and the system quickly attains a diagonal ensemble, and
$\chi(t)$ fluctuates about an average value $\chi(t
\gg \tau)\sim 1/L \sum_{m} |c_m|^2 \langle\phi_m|\hat
N_d|\phi_m\rangle$~\cite{rigol1,rigol2}, implying that the the peak
of $\tau(V_1)$ in Fig.~\ref{fig3}(d) is one-sided. This peak can be
used to detect the exceptional point.

\section{Discussion}
\label{diss} 

In this work, we have studied the Hatano-Nelson and the Su-Schrieffer-Heeger models
with non-reciprocal hops and nearest neighbor interactions. We have shown that these models provide
two examples of a class of non-Hermitian systems which have a
``phase'' with entirely real spectrum for sufficiently large interactions.

We note that none of these models have $\mathcal{PT}$ symmetry. Thus the generic expectation in the
existing literature for the existence of real eigenvalues in presence of such symmetry is not applicable
to the present models. Our work explicitly points out that the presence of global symmetries can not, by itself,
explain the reality of eigenspectrum of the model at large interaction strength.

In contrast, as we show in this work, such a phase with entirely real eigenspectrum is a consequence of two ingredients.
First, the dynamical constraints in the infinitely large interaction limit which also
fragments the Hilbert space of the models. This Hilbert space fragmentation leads to restriction
on the matrix elements of the Hamiltonian between Fock states and, as we show, allows for an existence of a many-body
similarity transformation which allows us to prove the reality of the eigenspectrum in this limit.
Second, moving away from the fragmented limit, we show that the presence of global symmetries allows
one to retain the reality of eigenspectrum up to a critical interaction strength. The effective
Hamiltonian between two states related by such symmetries (and which are therefore degenerate in the
limit of large $V$) can be shown to be Hermitian for finite $V \le V_c$; this allows for persistence of
real eigenvalues till a critical interaction strength is reached. Interestingly, this mechansim seems
to be effective even when such symmetries are weakly broken as seen from the analysis of the SSH model.
Thus our work points out the role of HSF for the presence of real eigenspectrum; to the best of our knowledge,
such application of HSF in the context of non-Hermitian system has not been pointed out earlier.

In addition we also provide a method of detection of the first exceptional point in these systems. Our analysis indicates that
time taken by an equal-time correlator of such a system to reach its steady state value following a quench diverges at the
first exceptional point. This observation, which only relies on the presence of an infinitesimal complex component of 
a pair of eigenvalues at this point, provides a link between presence of exceptional points and the behavior of experimentally
measurable equal-time correlation functions in such non-Hermitian systems.

To summarize, we reveal a deep link between the physics of fragmentation and the property of real
spectrum of an interacting non-Hermitian system. This link is worth
investigating in the future to identify and construct other systems belonging
to this class. We also propose a method to detect many-body exceptional points by studying the time
required for a correlation function to reach its steady state value.

\acknowledgements The authors thank Diptiman Sen for several
comments. IP is thankful to Masudul Haque for insightful
discussions. SG acknowledges the financial support provided by CSIR,
India through file 09/080(1133)/2019-EMR-I. KS thanks DST, India for
support through SERB project JCB/2021/000030.

\appendix

\section{Argument for fragmentation}
%%%%%%%%%%%%%%%%%%%%%%%%%%%%%%%%%%%%%%%%%%%%%
\begin{figure}
\includegraphics[width=1.0\linewidth]{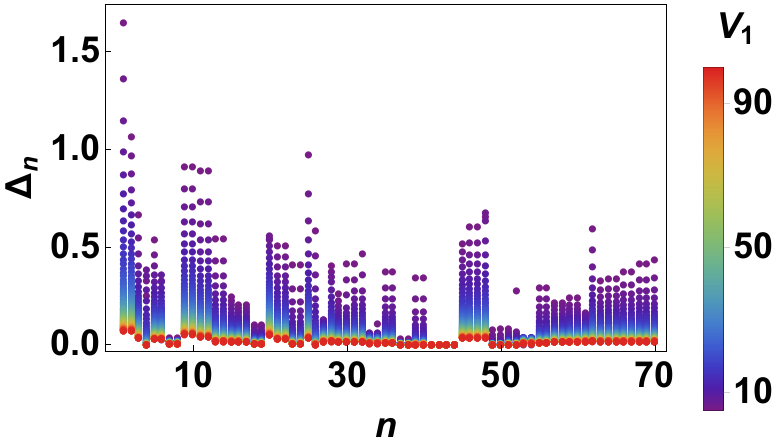}
\caption{Variation of the difference between the eigenvalues of $\ham$ and that of $\ham_{HN,f}$ with $V_1$ for $L=8$ with anti-PBC. As $V_1$ increases, $\ham_{HN,f}$ increasingly becomes a better approximation to $\ham_{HN}$. See text for details.} \label{sfig1}
\end{figure}
%%%%%%%%%%%%%%%%%%%%%%%%%%%%%%%%%%%%%%%%%%%%%%

In Fig. \ref{sfig1}, we plot the absolute difference, $\Delta_n=|\varepsilon_n-\varepsilon^f_n|$ between the energy eigenvalues $\varepsilon_n$ of the exact Hamiltonian $\ham_{HN}$ and $\varepsilon_n^f$ of the fragmented Hamiltonian $\ham_{HN,f}$. We take $L=8$ with anti-PBC. It can be seen that as the nearest neighbor interaction $V_1$ increases, the spectrum of $\ham_{HN}$ approaches that of $\ham_{HN,f}$. This implies that in the large $V_1$ limit, $\ham_{HN,f}$ gives the leading order description of $\ham_{HN}$.

Next, we revisit the argument why the constrained hopping of
$\ham_{HN,f}$ in Eq. 2 of the main text, leads to strong Hilbert space fragmentation.
For the Hermitian case the argument can be found in the literature in terms of
the mapping to ``spins and movers'' \cite{tomasi2019}. Since the argument is based in Fock space,
it is valid as well
for non reciprocal hopping. Below we reconstruct the argument in terms of ``particle-
and hole-defects''.

As already outlined in the main text, due to half-filling the defect locations follow
certain rules. (a)
If two particle-defects at $i_1$ and $i_2$ are ``adjacent'', then
$(i_1, i_2)$ can only be (odd, even) or (even, odd). The same
applies for two adjacent hole-defects. Two defects are
adjacent if there is no third defect in between the two while
traversing either clockwise or counterclockwise.
(b) If a particle-defect at $i_1$ is adjacent to a hole-defect at $j_1$,
then $(i_1) (j_1)$ can only be (even)(even) or (odd)(odd).

Also, the defect dynamics, due to the constrained hopping, obey the following rules.
(i) An allowed fermion hop changes $i$ or $j$ by $\pm$2 modulo $L$. (ii) Since
second nearest neighbor hopping is absent, two defects cannot cross each other.

Due to rule (ii) two configurations in which the sequence of the defects are not cyclically
related can never be connected by $\ham_{HN,f}$, and hence they belong to two different
fragments. Thus, each fragment can be labeled by the sequence in which the defects appear
while moving, say, from site 1 to $L$ of the chain. As an example consider the sector $N_d = 3$.
The wavefunctions have three particle-defects and three hole defects. Note, the sequences need
to obey rules (a) and (b) mentioned above. Moreover, due to rule (i), a defect on an odd (even)
site always stays on an odd (even) site.
There are eight possible different sequences that can
be designated as
\begin{eqnarray*}
  (p_e, h_e, p_e, h_e,p_e, h_e),\quad (p_e, h_e, p_e, p_o, h_o, h_e),\\
(p_e, p_o, p_e, h_e, h_o, h_e),\quad (p_o, p_e, h_e, p_e , h_e, h_o), \\
(p_o, h_o, p_o, p_e, h_e, h_o), \quad(p_0, p_e, p_o, h_o, h_e, h_o),\\
(p_e, p_o, h_o, p_o, h_o, h_e),\quad (p_o, h_o, p_o, h_o, p_o, h_o),
\end{eqnarray*}
where
$p_{o/e}$ implies a particle-defect on an odd (even) site, and $h_{o/e}$ implies a
hole-defect on an odd (even) site.
Thus, by suitably inserting hole defects, each of the three
particle defects can be put either on an odd or on an even site. This implies $2^{N_d}$
distinct fragments. In practice, in between a $p_e$ and a $p_o$ one can put an
even number, including zero, of hole-defects, and in between two $p_e$
one can put an odd number of hole-defects. This last flexibility makes the actual number of
fragments more than $2^{N_d}$ for $L$ and $N_d$ large enough. Thus, $2^{N_d}$ is a lower
bound, which nevertheless shows that the number of fragments proliferate exponentially with
the system size.

Next we enumerate the size of the largest fragment in the symmetry sector with $N_d = L/4$ defects.
Consider a seed state with all the particle-defects $p$ to the left, followed by all the hole-defects $h$
and then
pairs of filled and empty sites denoted by $R$ ($R\equiv \smblkcircle\smwhtcircle$). This configuration has the form
\[
(\underbrace{p, p, \ldots}_{N_d \, {\rm times}}, \underbrace{h, h, \ldots}_{N_d \, {\rm times}},
\underbrace{ R, R, \ldots}_{P \, {\rm times}}).
\]
Since $N_d$ consecutive particle-defects require $(N_d + 1)$ particles, and
$N_d$ consecutive hole-defects require $(N_d + 1)$ holes, while the remaining particle/holes are in pairs,
the number of pairs is $P = L/2 - (N_d + 1)$. Under the action of $\ham_{HN,f}$
a particle from $R$ can ``diffuse'' into the
sequence of hole-defects. However, it needs to be accompanied by a hole as well, in order to maintain
the number $N_d$ of hole-defects. The same is true for diffusion into the sequence of particle-defects. For instance,
\[ |\smblkcircle\smblkcircle\smblkcircle\smblkcircle\text{ }\smwhtcircle\text{ }\smwhtcircle\text{ }\smwhtcircle\text{ }\smwhtcircle\smblkcircle\smwhtcircle\smblkcircle\smwhtcircle\rangle \to |\smblkcircle\smblkcircle\smblkcircle\smblkcircle\text{ }\smwhtcircle\text{ }\smwhtcircle\text{ }\smwhtcircle\smblkcircle\smwhtcircle\text{ }\smwhtcircle\smblkcircle\smwhtcircle\rangle.
\]
In other words, only pairs diffuse under the constrained hopping. The
size of the fragment is given by the number of distinct wavefunctions that can be created from this seed state
through pair diffusion. Since $P$ pairs can be put in $2 N_d$ intermediate positions in between the defects,
and since translation of a given configuration by two sites also generates new configurations
(note, defects can move only by two sites), the size of
the largest fragment is
\[
N_f = \frac{L}{2} {{2N_d + P -1}\choose{P}} = \frac{L}{2} {{3L/4 -2}\choose{L/4 -1}}.
\]
On the other hand, the total size of the symmetry sector with $N_d$ defects is \cite{frey2022}
\[
N_s = 2 {{L/2}\choose{L/4}}{{L/2-1}\choose{L/4}}.
\]
In the limit of large chain length $L$, this implies that the ratio
$\lambda \equiv N_f/N_s \approx (0.8)^L$, which vanishes exponentially with the system size.
In other words, even the largest fragment samples only a vanishing portion of the total
possible states. This establishes strong Hilbert space fragmentation.

\section{More on Connectivity Diagrams and Similarity Transformation}

%%%%%%%%%%%%%%%%%%%%%%%%%%%%%%%%%%%%%%%%%%%%%%%%%%%%%%%
\begin{figure*}[t]
\includegraphics[width=0.8\linewidth]{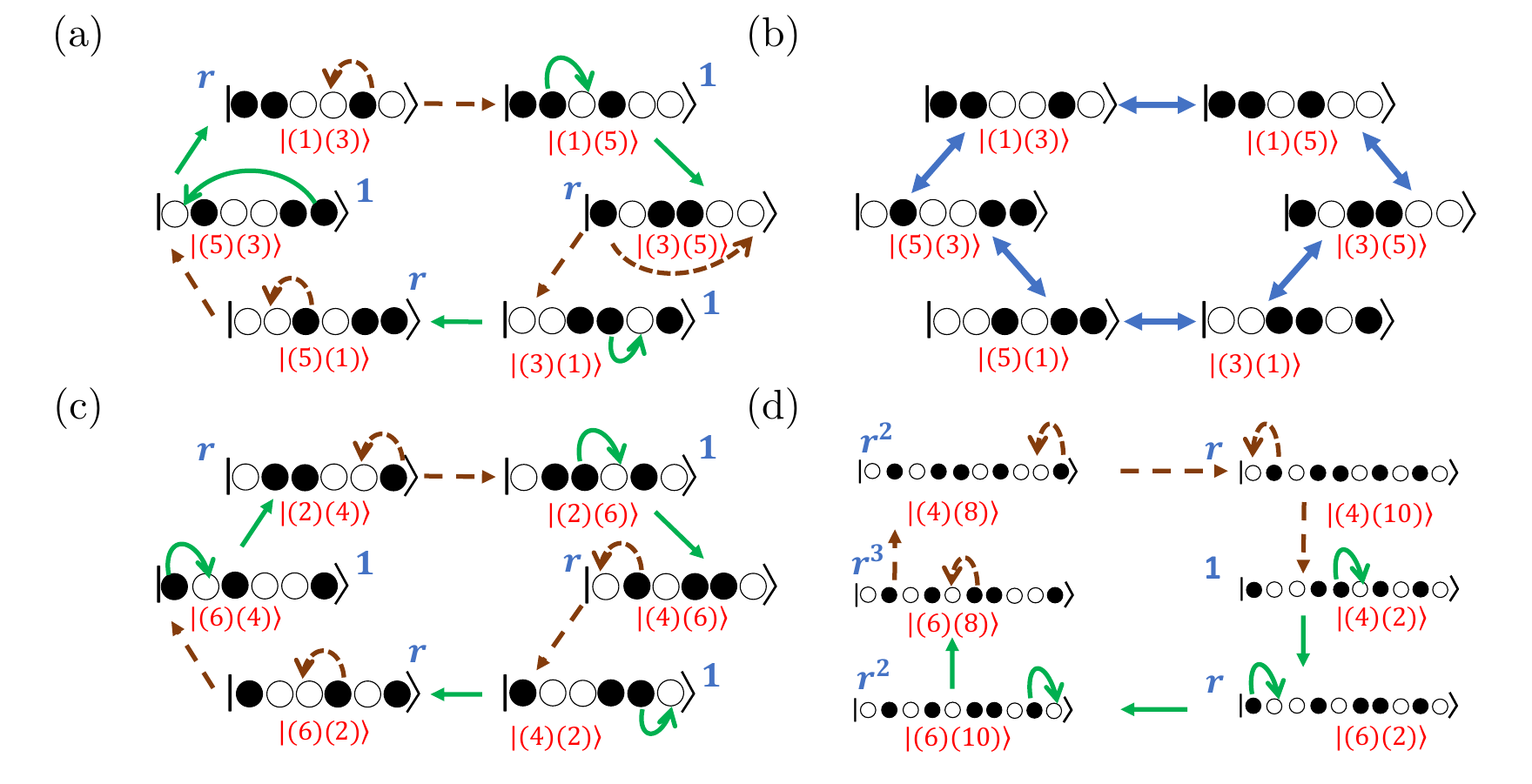}
\caption{(a) Connectivity diagram corresponding to $\ham_{HN,f}^{\text{odd}}$ showing both real space hoppings and the Fock space structure. Solid (green) and dashed (brown) arrows represent right hops with strength $J_2$ and left hops with strength $J_1$ respectively. The numbers in blue show the scaling factors of the states of the states, required for the similarity transformation $\mathcal{S}$. This gives an example of the second type of connectivity. (b) Connection diagram in Fock space of $\Tilde{\ham}_f^{\text{odd}}$ after the similarity transformation is carried out. Both the right and left hops, indicated by double-headed blue arrows, have strengths $\sqrt{J^2-\gamma^2}$ and hence the structure is hermitian. (c) Same as (a) but now for $\ham_{HN,f}^{\text{even}}$ in the (even)(even) sector. It can be seen that (c) is related to (a) by $\mathcal{R}$. After carrying out the similarity transformation $\mathcal{S}$, one gets a hermitian structure like that of (b) for the (even)(even) sector. (d) A closed loop from the $N_d=1$ sector of the half-filled $L=10$ chain in which the defects retrace back their paths to return to their initial position. This is an example of the first type of connectivity. (a)-(c) correspond to $N_d=1$ sector of a half-filled $L=6$ chain.} \label{sfig2}
\end{figure*}
%%%%%%%%%%%%%%%%%%%%%%%%%%%%%%%%%%%%%%%%%%%%%%%%%%%%%%%
\begin{figure*}[t]
\includegraphics[width=0.8\linewidth]{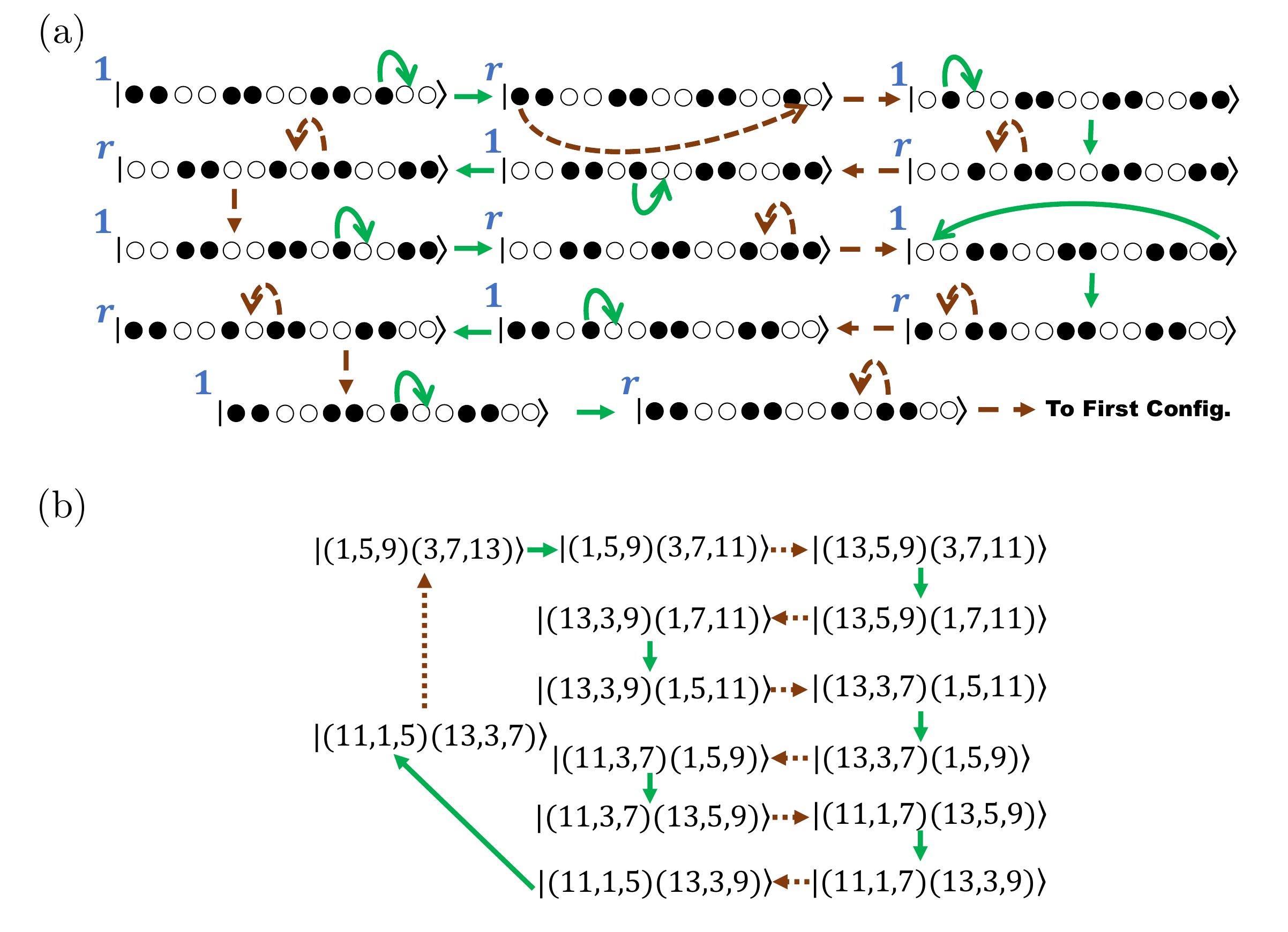}
\caption{(a) A closed loop from the $N_d=3$ sector of the half-filled $L=14$ chain where the defects permute their positions to reach the initial configuration. This is an example of the third kind of connectivity as specified in the main text. Solid (green) and dashed (brown) arrows represent right hops with strength $J_2$ and left hops with strength $J_1$ respectively. The numbers in blue show the scaling factors of the states of the states, required for the similarity transformation. (b) The same connectivity diagram in the Fock space with the states labelled according to the labelling scheme used in the main text.} \label{sfig3}
\end{figure*}
%%%%%%%%%%%%%%%%%%%%%%%%%%%%%%%%%%%%%%%%%%%%%%%%%%%%%%%%%

In this section, we provide an illustration of the similarity transformation in a simple case and explicit examples of the three possible types of closed loops as mentioned in the main text.

To begin with, we consider a chain of length $L=6$ having 3 fermions with PBC. We focus on the $N_d=1$ sector and since the Hamiltonian $\ham_{HN,f}$ changes the positions of the hole and the particle defects by $\pm 2$ (mod $L$) only, we consider the (odd)(odd) and the (even)(even) sectors separately. The basis states for the (odd)(odd) sector are enumerated as $\{|(1)(3)\rangle, |(1)(5)\rangle, |(3)(5)\rangle, |(3)(1)\rangle, |(5)(1)\rangle, |(5)(3)\rangle\}$. It's worthwhile to note that the labelling scheme for the states followed here are the same as outlined in the main text. In this basis, the Hamiltonian $\ham_{HN,f}^{\text{odd}}$ reads as
\begin{equation}
    \ham_{HN,f}^{\text{odd}}= \begin{pmatrix}
    V_1 & J_2 & 0 & 0 & 0 & J_2\\
    J_1 & V_1 & J_1 & 0 & 0 & 0\\
    0 & J_2 & V_1 & J_2 & 0 & 0\\
    0 & 0 & J_1 & V_1 & J_1 & 0\\
    0 & 0 & 0 & J_2 & V_1 & J_2\\
    J_1 & 0 & 0 & 0 & J_1 & V_1\end{pmatrix}
    \label{eqn:HFoddmat}
\end{equation}
where $J_1\equiv J-\gamma$ and $J_2\equiv J+\gamma$. The connectivity diagram corresponding to $\ham_{HN,f}^{\text{odd}}$ is shown in Fig. \ref{sfig2}(a). This is, in fact, an example of the second type of connectivity mentioned in the previous section wherein a particle defect moves across the entire chain to return to its original position. This structure is reminiscent of a single-particle hopping problem on a one-dimensional ring in Fock space but with different hopping strengths on alternate links. To map this to our familiar hopping problem, we define a similarity transformation
\begin{equation}
    \mathcal{S}= \begin{pmatrix}
    r & 0 & 0 & 0 & 0 & 0\\
    0 & 1 & 0 & 0 & 0 & 0\\
    0 & 0 & r & 0 & 0 & 0\\
    0 & 0 & 0 & 1 & 0 & 0\\
    0 & 0 & 0 & 0 & r & 0\\
    0 & 0 & 0 & 0 & 0 & 1\end{pmatrix}
    \label{eqn:Smat}
\end{equation}
with $r=\sqrt{J_2/J_1}$. The scalings are also shown in Fig. \ref{sfig2}(a) in blue. It's straightforward to see that under this transformation, $\ham_{HN,f}^{\text{odd}}$ assumes a hermitian structure given by
\begin{eqnarray}
\Tilde{\ham}_{HN,f}^{\text{odd}}&=&\mathcal{S}^{-1}\ham_{HN,f}^{\text{odd}}\mathcal{S}\nonumber\\
&=&\begin{pmatrix}
    V_1 & \alpha & 0 & 0 & 0 & \alpha\\
    \alpha & V_1 & \alpha & 0 & 0 & 0\\
    0 & \alpha & V_1 & \alpha & 0 & 0\\
    0 & 0 & \alpha & V_1 & \alpha & 0\\
    0 & 0 & 0 & \alpha & V_1 & \alpha\\
    \alpha & 0 & 0 & 0 & \alpha & V_1\end{pmatrix}
    \label{eqn:HF_effmat}
\end{eqnarray}
where $\alpha=\sqrt{J^2-\gamma^2}$.

The corresponding connectivity structure in Fock space is shown in Fig. \ref{sfig2}(b). This, of course, admits of a real spectrum and the eigenvalues can be readily written down as $\mathcal{E}_k = V_1 + 2\sqrt{J^2 - \gamma^2} \cos (k)$. Here the fictitious lattice spacing is taken to be unity and $k=-\pi + m\pi/3$, with $m=0,1,\ldots,5$. The eigenvectors $|k\rangle$ of the original problem can be recovered using the similarity transformation $\mathcal{S}$.

It's useful to note here that the (even)(even) sector is related to the (odd)(odd) sector through a translation by one lattice spacing $\mathcal{R}$. This is evident on comparing the connection diagrams in Fig. \ref{sfig2}(a) and (c). Hence $\ham_{HN,f}^{\text{even}}$ has exactly the same structure and spectrum as that of $\ham_{HN,f}^{\text{odd}}$. Equivalently, $\ham_{HN,f}^{N_d=1} =\ham_{HN,f}^{\text{odd}} \oplus \ham_{HN,f}^{\text{even}}$ is invariant under $\mathcal{R}$. This property would be significant in the next section.

Before closing this section, we chalk out examples of the other types of connectivities discussed in the main text. Fig. \ref{sfig2}(d) illustrates the situation where a particle defect completes a loop by retracing back its path. This gives an example of the first type of connectivity discussed in the main text. Fig. \ref{sfig3} shows a case where multiple defects exchange their positions in order to complete the loop. This is an example of the third type of connectivity. It's worth noting that in all these three cases, there are an equal number of right hops and left hops, which makes the similarity transformation possible. As the system size increases, more complicated connection diagrams emerge. However, any closed loop in these connection diagrams will fall in one of these three classes or will be some combination of these three and hence will admit of this general feature.\\

\section{Symmetry Protection and Hidden Hermiticity}

We use this section to illustrate with a concrete example, the symmetry protection of the real eigenvalues
once we move out of the limit of fragmentation of $\ham_{HN,f}$ (i.e., infinite $V_1$), and consider large but finite $V_1$ effects.

We again consider the $N_d=1$ sector of the half-filled chain of length $L=6$, whose spectrum was discussed in the previous section. Corresponding to the eigenvalue $\mathcal{E}_{\pi/3}$, there would be 4 degenerate states of $\ham_{HN,f}$, namely $|\psi_{1,2}\rangle_o$ and $|\psi_{1,2}\rangle_e$.
\begin{eqnarray*}
    |\psi_1\rangle_o &=& \frac{1}{\sqrt{3}} \big(r \mathcal{C} |(1)(3)\rangle - \mathcal{C} |(1)(5)\rangle - r |(3)(5)\rangle -\mathcal{C} |(3)(1)\rangle\nonumber\\
    &+& r \mathcal{C} |(5)(1)\rangle+ |(5)(3)\rangle \big) \nonumber\\
    |\psi_2\rangle_o &=& \frac{1}{\sqrt{3}} \mathcal{S} \big( r |(1)(3)\rangle +|(1)(5)\rangle -|(3)(1)\rangle - r |(5)(1)\rangle\big) \nonumber\\
\end{eqnarray*}
\begin{eqnarray*}
    |\psi_1\rangle_e &=& \frac{1}{\sqrt{3}} \big(r \mathcal{C} |(2)(4)\rangle - \mathcal{C} |(2)(6)\rangle - r |(4)(6)\rangle -\mathcal{C} |(4)(2)\rangle\nonumber\\
    &+& r \mathcal{C} |(6)(2)\rangle+ |(6)(4)\rangle \big) \nonumber\\
    |\psi_2\rangle_e &=& \frac{1}{\sqrt{3}} \mathcal{S} \big( r |(2)(4)\rangle +|(2)(6)\rangle -|(4)(2)\rangle - r |(6)(2)\rangle\big) \nonumber\\
\end{eqnarray*}
where $r = (J_2/J_1)^{1/2}$,
$\mathcal{C}=\cos{\frac{\pi}{3}}$ and $\mathcal{S}=\sin{\frac{\pi}{3}}$.

The first two states are from the (odd)(odd) sector and lie in the subspace spanned by $\big|k=\pm \frac{\pi}{3}\rangle$, which are related by $\mathcal{P}\mathcal{C}$ symmetry. This can be seen easily considering that under $\mathcal{P}\mathcal{C}$, $\{|(1)(3)\rangle \leftrightarrow |(3)(5)\rangle, |(3)(1)\rangle \leftrightarrow |(5)(3)\rangle, |(1)(5)\rangle \leftrightarrow |(1)(5)\rangle, |(5)(1)\rangle \leftrightarrow |(5)(1)\rangle\}$. The same can be said about the last two states which lie in the (even)(even) sector. States from the (odd)(odd) sector and (even)(even) sector are related by $\mathcal{R}$, viz $|\psi_{1(2)}\rangle_e = \mathcal{R} |\psi_{1(2)}\rangle_o$. It is useful to note here that in addition $(\mathcal{P}\mathcal{C})^2=\mathcal{I}$ and $\mathcal{R}^2 |\psi_{1(2)}\rangle_{o,e}=|\psi_{1(2)}\rangle_{o,e}$.

We consider the matrix representation of the second order correction, $\ham_{HN}^{(2)}$ in this degenerate subspace. Enumerating the basis as $\{|\psi_1\rangle_o, |\psi_2\rangle_o, |\psi_1\rangle_e, |\psi_2\rangle_e\}$, $\mathcal{H_{HN}}^{(2)}$ reads
\begin{equation}
    \ham_{HN}^{(2)}=\begin{pmatrix}
        \mu & 0 & \nu & \lambda \\
        0 & \mu & -\lambda & \nu\\
        \nu & -\lambda & \mu & 0\\
        \lambda & \nu & 0 & \mu
    \end{pmatrix}
    \label{eq:H2}
\end{equation}
where $\mu=-\frac{J_1 J_2}{V_1}(1+r^2)$, $\nu=-\frac{r}{4V_1} \left( J_1^2 +J_2^2\right)$ and $\lambda=\frac{\sqrt{3}}{4V_1}r\left( J_1^2 +J_2^2\right)$. This is Hermitian and admits of real spectrum. This hermiticity is expected to hold order-by-order in perturbation since Eq. \ref{eq:symprot} in the min text is true for exact $\ham_{HN}$. Thus, the reality of the spectrum is guaranteed within this subspace. %Any higher order corrections would be at most of order $(J,\gamma)^3/V_1^2$ and hence will not destroy the reality of the spectrum in this subspace.

There remains a second possibility in which these degeneracies might be accidental, i.e. not related by any symmetry, at some point in parameter space. However, we would argue here that such accidental degeneracies are not possible in generic integrable or non-integrable systems. In case of generic finite-sized non-integrable systems, one generally expects level repulsion leading to finite gaps in the spectrum, which protects the reality of the eigenspectrum at finite $V_1$. These gaps decrease with increasing size as pointed out in the main text; consequently, we expect $V_c$ to diverge in the thermodynamic limit. And in case of an integrable model, a generic crossing between two eigenstates involves eigenstates from different conserved charge sectors.
The full Hamiltonian, being integrable, cannot connect between states in different sectors at any order in $1/V_1$. Thus, there is no matrix element between them once $V_1$
is lowered. For our case, we have not observed any occurrence of accidental degeneracy.

\section{Boundary Condition, Symmetry Breaking and Complex Eigenvalues}
%%%%%%%%%%%%%%%%%%%%%%%%%%%%%%%%%%%%%%%%%%%%%%%%%%%
\begin{figure}
\includegraphics[width=0.8\linewidth]{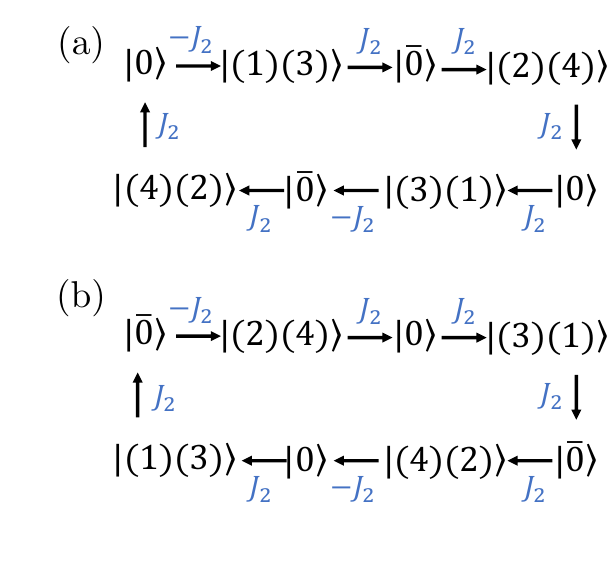}
\caption{(a) Fock space connectivity diagram corresponding to the full Hamiltonian $\ham_{HN}$ for $L=4$ with periodic boundary condition. Letters in blue above the arrows show the hopping strength for that particular hop. (b) The fate of the connectivity diagram on application of the one-site lattice translation operator $\mathcal{R}$. Comparison of the two figures clearly show that $\ham_{HN}$ with this boundary condition is not invariant under $\mathcal{R}$. See text for details.}
\label{sfig4}
\end{figure}
%%%%%%%%%%%%%%%%%%%%%%%%%%%%%%%%%%%%%%%%%%%%%%%%%%%

So far, we have considered PBC for odd filling and anti-PBC for even filling. In this section, we show that choosing the opposite set of boundary conditions breaks the translational invariance of the full Hamiltonian $\ham_{HN}$ and results in complex eigenvalues for any finite value of $V_1$. This happens even though $\ham_{HN,f}$ has real eigenvalues just as in the case of the standard boundary condition. This is an example where there is no sufficient symmetry protection in the degenerate subspace of $\ham_{HN,f}$, and therefore $V_c$ is trivially infinite, because any finite $V_1$ will lead to complex eigenvalues.

For illustration purposes, we consider a half-filled chain of length $L=4$ with PBC. Choosing the basis as $\{|0\rangle,|\Bar{0}\rangle,|(1)(3)\rangle,|(2)(4)\rangle,|(3)(1)\rangle,|(4)(2)\rangle\}$, the Hamiltonian $\ham_{HN}$ can be written as
\begin{equation}
    \ham_{HN}=\begin{pmatrix}
        0 & 0 & -J_1 & J_2 & J_1 & J_2\\
        0 & 0 & J_2 & J_1 & -J_2 & J_1\\
        -J_2 &  J_1 & V_1 & 0 & 0 & 0\\
        J_1 &  J_2 & 0 & V_1 & 0 & 0\\
        J_2 &  -J_1 & 0 & 0 & V_1 & 0\\
        J_1 &  J_2 & 0 & 0 & 0 & V_1\\
    \end{pmatrix}
    \label{eq:Hpbc4}
\end{equation}
where $|0\rangle \equiv |\smwhtcircle\smblkcircle\smwhtcircle\text{ }\smblkcircle\rangle$ and $|\Bar{0}\rangle \equiv |\smblkcircle\smwhtcircle\smblkcircle\smwhtcircle\rangle$. The corresponding connection diagram is shown in Fig. \ref{sfig4}(a). It can be readily verified from the matrix structure as well as from the connection diagram that $\ham$ is not invariant under $\mathcal{R}$ which maps $|0\rangle\leftrightarrow|\Bar{0}\rangle \text{ and } |(1)(3)\rangle\to |(2)(4)\rangle\to|(3)(1)\rangle\to|(4)(2)\rangle\to|(1)(3)\rangle$.

However, $\ham_{HN,f}$ in this basis reads
\begin{equation}
 \ham_{HN,f}=\begin{pmatrix}
        0 & 0 & 0 & 0 & 0 & 0\\
        0 & 0 & 0 & 0 & 0 & 0\\
        0 & 0 & V_1 & 0 & 0 & 0\\
        0 & 0 & 0 & V_1 & 0 & 0\\
        0 & 0 & 0 & 0 & V_1 & 0\\
        0 & 0 & 0 & 0 & 0 & V_1\\
    \end{pmatrix}
    \label{eq:Hfpbc4}
\end{equation}
This is trivially Hermitian in this case because all the basis states are annihilated by the constrained hopping terms in $\ham_{HN,f}$. Such states are termed as frozen states in the field of fragmentation physics. The energy eigenstates can be chosen to be the basis states and they are doubly degenerate in the $N_d=0$ sector and exhibit 4-fold degeneracy in the $N_d=1$ sector.

We now consider the representation of the second-order correction to $\ham_{HN}$ in both these degenerate sub-spaces.
\begin{equation}
    \ham_{HN}^{(2)}=\begin{pmatrix}
        4\mu & -2\delta & 0 & 0 & 0 & 0\\
        2\delta & 4\mu & 0 & 0 & 0 & 0\\
        0 & 0 & -2\mu & -\delta & 2\mu & -\delta\\
        0 & 0 & \delta & -2\mu & -\delta & -2\mu\\
        0 & 0 & 2\mu & \delta & -2\mu & \delta\\
        0 & 0 & \delta & -2\mu & -\delta & -2\mu
    \end{pmatrix}
    \label{eq:H2pbc4}
\end{equation}
where $\mu=-\frac{J_1J_2}{V_1}$ and $\delta=\frac{1}{V_1}\left(J_2^2-J_1^2\right)$. This is non-Hermitian, implying that for any finite value of $V_1$, the eigenspectrum will be complex. Explicitly, the eigenvalues of $\ham_{HN,f}+\ham_{HN}^{(2)}$ are
\begin{equation}
    \epsilon=\left(V_1,V_1,-\frac{4}{V_1}(J\pm i\gamma)^2, V_1+\frac{4}{V_1}(J\pm i\gamma)^2\right)
    \label{eq:eigvalwrongbc}
\end{equation}
This makes it impossible to have a real eigenspectrum for any finite $V_1$; this fact has been
numerically found in Ref.\ \cite{zhang2022}.

It is important to note here that $\ham_{HN,f}$ yields a real spectrum irrespective of the chosen boundary condition. In the case shown, it is trivially real; in other cases, where the connection diagrams are more complicated, a similarity transformation like the above will smoothly go through giving rise to a real spectrum. The hidden Hermiticity
argument becomes invalid because Eq. \ref{eq:symprot} of the main text fails,
owing to the lack of translational symmetry $\mathcal{R}$ of the full
Hamiltonian $\ham_{HN}$. This results in the higher-order corrections to be manifestly non-Hermitian in the degenerate sub-spaces, thereby immediately destroying the reality of the spectrum for any finite $V_1$. \\


\begin{thebibliography}{99}

\bibitem{reviews_nonH}
For reviews see, e.g., I. Rotter, J. Phys. A: Math. Theor. {\bf 42},
153001 (2009); Y. Ashida, Z. Gong, and M. Ueda, Adv. Phys. {\bf 69},
249 (2021); I. Rotter and J. P. Bird,  Rep. Prog. Phys. {\bf 78},
114001 (2015).

\bibitem{nonhlit1}J. Gonzalez and R. A. Molina, Phys. Rev. B {\bf 96}, 045437
(2017); V. Kozii and L. Fu, arXiv:1708.05841 (unpublished); A. A.
Zyuzin and A. Y. Zyuzin, Phys. Rev. B {\bf 97}, 041203(R) (2018); H.
Shen and L. Fu, Phys. Rev. Lett. {\bf 121}, 026403 (2018); R. A.
Molina and J. Gonzalez, Phys. Rev. Lett. {\bf 120}, 146601 (2018);T.
Yoshida, R. Peters, and N. Kawakami, Phys. Rev. B {\bf 98}, 035141
(2018); J. Carlstrom and E. J. Bergholtz, Phys. Rev. A {\bf 98},
042114 (2018).

\bibitem{nonhlit2} T. M. Philip, M. R. Hirsbrunner, and M. J. Gilbert, Phys. Rev. B
{\bf 98}, 155430 (2018); Y. Chen and H. Zhai, Phys. Rev. B {\bf 98},
245130 (2018); K. Moors, A. A. Zyuzin, A. Y. Zyuzin, R. P. Tiwari,
and T. L. Schmidt, Phys. Rev. B {\bf 99}, 041116(R) (2018); R.
Okugawa and T. Yokoyama, Phys. Rev. B {\bf 99}, 041202(R) (2019); J.
C. Budich, J. Carlstrom, F. K. Kunst, and E. J. Bergholtz, Phys.
Rev. B {\bf 99}, 041406(R) (2019).

\bibitem{nonhlit3} Z. Yang and J. Hu, Phys. Rev. B {\bf 99}, 081102(R) (2019); T. Yoshida,
R. Peters, N. Kawakami, and Y. Hatsugai, Phys. Rev. B {\bf 99},
121101(R) (2019); Y. Wu, W. Liu, J. Geng, X. Song, X. Ye, C.-K.
Duan, X. Rong, and J. Du, Science {\bf 364}, 878 (2019); P.
San-Jose, J. Cayao, E. Prada, and R. Aguado, Sci. Rep. {\bf 6},
21427 (2016); Q.-B. Zeng, B. Zhu, S. Chen, L. You, and R. Lu, Phys.
Rev. A {\bf 94}, 022119 (2016); C. Li, X. Z. Zhang, G. Zhang, and Z.
Song, Phys. Rev. B {\bf 97}, 115436 (2018); J. Cayao and A. M.
Black-Schaffer Phys. Rev. B {\bf 105}, 094502 (2022); R. Arouca, J.
Cayao, A. M. Black-Schaffer, arXiv:2206.15324 (unpublished).

\bibitem{nonhlit4} K. Kawabata, Y. Ashida, H. Katsura, and M. Ueda, Phys. Rev. B {\bf 98},
085116 (2018); A. Guo, G. J. Salamo, D. Duchesne, R. Morandotti, M.
Volatier-Ravat, V. Aimez, G. A. Siviloglou, and D. N.
Christodoulides, Phys. Rev. Lett. {\bf 103}, 093902 (2009);  C. E.
Ruter, K. G. Makris, R. El-Ganainy, D. N. Christodoulides, M. Segev,
and D. Kip, Nat. Phys. {\bf 6}, 192 (2010); L. Feng, M. Ayache, J.
Huang, Y.-L. Xu, M.-H. Lu, Y.-F. Chen, Y. Fainman, and A. Scherer,
Science 333, 729 (2011); A. Regensburger, C. Bersch, M.-A. Miri, G.
Onishchukov, D. N. Christodoulides, and U. Peschel, Nature (London)
{\bf 488}, 167 (2012).

\bibitem{nonhlit5} L. Feng, Y.-L. Xu, W. S. Fegadolli, M.-H. Lu, J. E. Oliveira, V.
R. Almeida, Y.-F. Chen, and A. Scherer, Nat. Mater. {\bf 12}, 108
(2013); C. Poli, M. Bellec, U. Kuhl, F. Mortessagne, and H.
Schomerus, Nat. Commun. {\bf 6}, 6710 (2015); B. Zhen, C.W. Hsu, Y.
Igarashi, L. Lu, I. Kaminer, A. Pick, S.-L. Chua, J. D.
Joannopoulos, and M. Solja.i., Nature (London) {\bf 525}, 354
(2015); H. Zhao, S. Longhi, and L. Feng, Sci. Rep. {\bf 5}, 17022
(2015); K. Ding, Z. Q. Zhang, and C. T. Chan, Phys. Rev. B {\bf 92},
235310 (2015).

\bibitem{nonhlit6} S. Weimann, M. Kremer, Y. Plotnik, Y. Lumer, S. Nolte, K. Makris,
M. Segev, M. Rechtsman, and A. Szameit, Nat. Mater. {\bf 16}, 433
(2017); H. Hodaei, A. U. Hassan, S. Wittek, H. Garcia-Gracia, R.
El-Ganainy, D. N. Christodoulides, and M. Khajavikhan, Nature
(London) {\bf 548}, 187 (2017); W. Chen, K. Ozdemir, G. Zhao, J.
Wiersig, and L. Yang, Nature (London) {\bf 548}, 192 (2017); P.
St-Jean, V. Goblot, E. Galopin, A. Lemaitre, T. Ozawa, L. Le
Gratiet, I. Sagnes, J. Bloch, and A. Amo, Nat. Photonics {\bf 11},
651 (2017).

\bibitem{nonhlit7} B. Bahari, A. Ndao, F. Vallini, A. E. Amili, Y. Fainman,
and B. K. Le, Science {\bf 358}, 636 (2017); J. Wang, H. Y. Dong, Q.
Y. Shi, W. Wang, and K. H. Fung, Phys. Rev. B {\bf 97}, 014428
(2018); H. Zhou, C. Peng, Y. Yoon, C.W. Hsu, K. A. Nelson, L. Fu, J.
D. Joannopoulos, M. Solja.i., and B. Zhen, Science {\bf 359}, 1009
(2018); M. Parto, S.Wittek, H. Hodaei, G. Harari, M. A. Bandres, J.
Ren, M. C. Rechtsman, M. Segev, D. N. Christodoulides, and M.
Khajavikhan, Phys. Rev. Lett. {\bf 120}, 113901 (2018); H. Zhao, P.
Miao, M. H. Teimourpour, S. Malzard, R. El-Ganainy, H. Schomerus,
and L. Feng, Nat. Commun. {\bf 9}, 981 (2018).

\bibitem{nonhlit8} G. Harari, M. A. Bandres, Y. Lumer, M. C. Rechtsman, Y.
D. Chong, M. Khajavikhan, D. N. Christodoulides, and M. Segev,
Science {\bf 359}, 1230 (2018); M. A. Bandres, S. Wittek, G. Harari,
M. Parto, J. Ren, M. Segev, D. N. Christodoulides, and M.
Khajavikhan, Science {\bf 359}, 1231 (2018); M. Pan, H. Zhao, P.
Miao, S. Longhi, and L. Feng, Nat. Commun. {\bf 9}, 1308 (2018); L.
Jin and Z. Song, Phys. Rev. Lett. 121, 073901 (2018); S. Malzard and
H. Schomerus, Phys. Rev. A {\bf 98}, 033807 (2018); Z. Oztas and C.
Yuce, Phys. Rev. A {\bf 98}, 042104 (2018).

\bibitem{nonhlit9} M. Kremer, T. Biesenthal, L. J. Maczewsky, M. Heinrich, R.
Thomale, and A. Szameit, Nat. Commun. {\bf 10}, 435 (2019); K. Y.
Bliokh, D. Leykam, M. Lein, and F. Nori, Nat. Commun. {\bf 10}, 580
(2019);  S. Wang, B. Hou, W. Lu, Y. Chen, Z. Zhang, and C. Chan,
Nat. Commun. {\bf 10}, 832 (2019); S. Chen,W. Zhang, B. Yang, T.Wu,
and X. Zhang, Sci. Rep. {\bf 9}, 5551 (2019); T. E. Lee and C.-K.
Chan, Phys. Rev. X {\bf 4}, 041001 (2014); Y. Xu, S.-T. Wang, and
L.-M. Duan, Phys. Rev. Lett. {\bf 118}, 045701 (2017); Y. Ashida, S.
Furukawa, and M. Ueda, Nat. Commun. {\bf 8}, 15791 (2017); Z. Gong,
Y. Ashida, K. Kawabata, K. Takasan, S. Higashikawa, and M. Ueda,
Phys. Rev. X {\bf 8}, 031079 (2018); M. Nakagawa, N. Kawakami, and
M. Ueda, Phys. Rev. Lett. {\bf 121}, 203001 (2018); K. Takata and M.
Notomi, Phys. Rev. Lett. {\bf 121}, 213902 (2018); L. Pan, S. Chen,
and X. Cui, Phys. Rev. A {\bf 99}, 011601(R) (2019).

\bibitem{nonhlit10} J. Li, A. K. Harter, J. Liu, L. de Melo, Y. N. Joglekar, and L.
Luo, Nat. Commun. {\bf 10}, 855 (2019); T. Liu, Y.-R. Zhang, Q. Ai,
Z. Gong, K. Kawabata, M. Ueda, and F. Nori, Phys. Rev. Lett. {\bf
122}, 076801 (2019); M. S. Rudner and L. S. Levitov, Phys. Rev.
Lett. {\bf 102}, 065703 (2009); J. M. Zeuner, M. C. Rechtsman, Y.
Plotnik, Y. Lumer, S. Nolte, M. S. Rudner, M. Segev, and A. Szameit,
Phys. Rev. Lett. {\bf 115}, 040402 (2015); K. Mochizuki, D. Kim, and
H. Obuse, Phys. Rev. A {\bf 93}, 062116 (2016); L. Xiao, X. Zhan, Z.
Bian, K. Wang, X. Zhang, X. Wang, J. Li, K. Mochizuki, D. Kim, N.
Kawakami et al., Nat. Phys. {\bf 13}, 1117 (2017).

\bibitem{nonhlit11} N. Hatano and D. R. Nelson, Phys. Rev. Lett. {\bf 77}, 570 (1996); N.
Hatano and D. R. Nelson, Phys. Rev. B {\bf 56}, 8651 (1997); N.
Hatano and D. R. Nelson, Phys. Rev. B {\bf 58}, 8384 (1998).

\bibitem{nonhlit12} J. A. S. Lourenco, R. L. Eneias, and R. G. Pereira, Phys. Rev. B
{\bf 98}, 085126 (2018); E. I. Rosenthal, N. K. Ehrlich, M. S.
Rudner, A. P. Higginbotham, and K.W. Lehnert, Phys. Rev. B {\bf 97},
220301(R) (2018); M. Wang, L. Ye, J. Christensen, and Z. Liu, Phys.
Rev. Lett. {\bf 120}, 246601 (2018).

\bibitem{nonhlit13} M. Ezawa, Phys. Rev. B {\bf 99}, 121411(R) (2019); M. Ezawa, Phys. Rev.
B {\bf 99}, 201411(R) (2019); M. Ezawa, Phys. Rev. B {\bf 100},
045407 (2019).


\bibitem{nhdyn1} L. Zhou, Q.-h.Wang, H.Wang, and J. Gong, Phys. Rev. A {\bf 98},
022129 (2018); L. Zhou and Q. Du, New J. Phys. {\bf 23}, 063041
(2021); B. Zhu, Y. Ke, H. Zhong, and C. Lee, Phys. Rev. Research
{\bf 2}, 023043 (2020); L. Zhou and J. Gong, Phys. Rev. B {\bf 98},
205417 (2018); L. Zhou, Phys. Rev. B {\bf 100}, 184314 (2019); L.
Zhou, Y. Gu, and J. Gong, Phys. Rev. B {\bf 103}, L041404 (2021).

\bibitem{nhdyn2} L. Zhou and W. Han, Phys. Rev. B {\bf 106}, 054307 (2022); C-H Liu, H.
Hu, and S. Chen, Phys. Rev. B {\bf 105}, 214305 (2022); L. Zhou, R.
W. Bomantara, and S. Wu, SciPost Phys. {\bf 13}, 015 (2022).

\bibitem{nhdyn3} S. Zamani, R. Jafari, and A. Langari, Phys. Rev. B {\bf 102}, 144306
(2020); R. Jafari and A. Akbari, Phys. Rev. A {\bf 103}, 012204
(2021); K. Yang, L. Zhou,W. Ma, X. Kong, P.Wang, X. Qin, X. Rong,
Y.Wang, F. Shi, J. Gong, and J. Du, Phys. Rev. B {\bf 100}, 085308
(2019); D. Chowdhury, A. Banerjee, and A. Narayan Phys. Rev. A {\bf
103}, L051101 (2021).

\bibitem{nhdyn4}P. He and Z-H Huang, Phys. Rev. A {\bf 102}, 062201
(2020); S. Longhi, J. Phys. A: Math. Theor. {\bf 50}, 505201 (2017).

\bibitem{nhdyn5} X. Turkeshi and M. Schiro, arXiv:2201.09895
(unpublished); T. Banerjee and K. Sengupta, Phys. Rev. B {\bf 107},
155117 (2023).


\bibitem{nhdyn6} J. Ren, P. Hanggi, and B. Li,
Phys. Rev. Lett. {\bf 104}, 170601 (2010); J. Ren, S. Liu, and B. Li
Phys. Rev. Lett. {\bf 108}, 210603 (2012); H. Xu, D. Mason, L. Jiang
and J. G. E. Harris, Nature {\bf 537}, 80 (2016); Z. Wang, J. Chen,
and J. Ren, Phys. Rev. E {\bf 106}, L032102 (2020);L. J.
Fernández-Alcazar, R. Kononchuk, H. Li, and T. Kottos, Phys. Rev.
Lett. {\bf 126}, 204101 (2021).

\bibitem{reviews_EP}
for reviews see, e.g., E. J. Bergholtz, J. C. Budich, and F. K. Kunst,
Rev. Mod. Phys. {\bf 93}, 015005 (2021);
W. D. Heiss, J. Phys. A: Math. Theor. {\bf 45}, 444016 (2012);
M. M\"{u}ller and I. Rotter, J. Phys. A: Math. Theor. {\bf 41}, 244018 (2008).


\bibitem{eptop1} Y. C. Hu and T. L. Hughes, Phys. Rev. B {\bf  84}, 153101 (2011); K. Esaki,
M. Sato, K. Hasebe, and M. Kohmoto, Phys. Rev. B 84, 205128 (2011);
T. E. Lee, Phys. Rev. Lett. {\bf 116}, 133903 (2016); D. Leykam, K.
Y. Bliokh, C. Huang, Y. D. Chong, and F. Nori, Phys. Rev. Lett. {\bf
118}, 040401 (2017); V. M. Martinez Alvarez, J. E. Barrios Vargas,
and L. E. F. Foa Torres, Phys. Rev. B {\bf 97}, 121401(R) (2018); Y.
Xiong, J. Phys. Commun. {\bf 2}, 035043 (2018); H. Shen, B. Zhen,
and L. Fu, Phys. Rev. Lett. {\bf 120}, 146402 (2018).

\bibitem{eptop2} C. Yuce, Phys. Rev. A {\bf 97}, 042118 (2018); C. Yin, H. Jiang, L. Li,
R. Lu, and S. Chen, Phys. Rev. A {\bf 97}, 052115 (2018); C. Yuce,
Phys. Rev. A {\bf 98}, 012111 (2018); F. K. Kunst, E. Edvardsson, J.
C. Budich, and E. J. Bergholtz, Phys. Rev. Lett. {\bf 121}, 026808
(2018); S. Yao, F. Song, and Z. Wang, Phys. Rev. Lett. {\bf 121},
136802 (2018); K. Kawabata, K. Shiozaki, and M. Ueda, Phys. Rev. B
{\bf 98}, 165148 (2018); C. Yuce and Z. Oztas, Sci. Rep. {\bf 8},
17416 (2018).

\bibitem{eptop3} S. Yao and Z. Wang, Phys. Rev. Lett. {\bf 121}, 086803 (2018)

\bibitem{eptop4} K. Kawabata, S. Higashikawa, Z. Gong, Y. Ashida, and M. Ueda,
Nat. Commun. {\bf 10}, 297 (2019); L. Jin and Z. Song, Phys. Rev. B
{\bf 99}, 081103(R) (2019); H. Wang, J. Ruan, and H. Zhang, Phys.
Rev. B {\bf 99}, 075130 (2019); D. S. Borgnia, A. J. Kruchkov, and
R.-J. Slager, Phys. Rev. Lett. {\bf 124}, 056802 (2020); Z.
Ozcakmakli Turker and C. Yuce, Phys. Rev. A {\bf 99}, 022127 (2019);
E. Edvardsson, F. K. Kunst, and E. J. Bergholtz, Phys. Rev. B {\bf
99}, 081302(R) (2019).

\bibitem{eptop5} C.-H. Liu, H. Jiang, and S. Chen, Phys. Rev. B {\bf 99}, 125103 (2019);
C. H. Lee and R. Thomale, Phys. Rev. B {\bf 99}, 201103(R) (2019);
F. K. Kunst and V. Dwivedi, Phys. Rev. B {\bf 99}, 245116 (2019); K.
Yokomizo and S. Murakami, Phys. Rev. Lett. {\bf 123} 066404 (2019).

\bibitem{eptop6} R. Nehra, and D. Roy, Phys. Rev. B {\bf 105}, 195407
(2022); K. Kawabata, K. Shiozaki, and S. Ryu, Phys. Rev. B {\bf
105}, 165137 (2022); K. Yang, D. Varjas, E. J. Bergholtz, S.
Morampudi, and F. Wilczek, arXiv:2202.04435 (unpublished).

\bibitem{Mostafazadeh2002a}
A. Mostafazadeh, J. Math. Phys. {\bf 43}, 205 (2002).

\bibitem{Mostafazadeh2002b}
A. Mostafazadeh, J. Math. Phys. {\bf 43}, 2814 (2002).

\bibitem{bender2007}
C. M Bender Rep. Prog. Phys. {\bf 70},  947 (2007).

\bibitem{zyablovsky2014}
A. A. Zyablovsky, A. P. Vinogradov, A. A. Pukhov, A. V. Dorofeenko, A. A. Lisyansky,
Phys.-Uspekhi {\bf 57}, 1063 (2014).

\bibitem{ozdemir2019}
S. K. \"{O}zdemir, S. Rotter, F. Nori, and L. Yang, Nat. Mater. {\bf 18}, 783 (2019).

%\bibitem{hatano1996}
%N. Hatano and D. R. Nelson, Phys. Rev. Lett. {\bf 77}, 570 (1996).
\bibitem{khemani2020}
V. Khemani, M. Hermele and R. Nandkishore, Phys. Rev. B {\bf 101},
174204 (2020).

\bibitem{sala2020}
P. Sala, T. Rakovszky, R. Verresen, M. Knap and F. Pollmann,
Phys. Rev. X {\bf 10}, 011047 (2020).

\bibitem{rakovszky2020}
T. Rakovszky, P. Sala, R. Verresen, M. Knap and F. Pollmann,
Phys. Rev. B {\bf 101}, 125126 (2020).

\bibitem{yang2020}
Z.-C. Yang, F. Liu, A. V. Gorshkov and T. Iadecola,
Phys. Rev. Lett. {\bf 124}, 207602 (2020).

\bibitem{tomasi2019}
G. De Tomasi, D. Hetterich, P. Sala,
and F. Pollmann, Phys. Rev. B {\bf 100}, 214313(2019).

\bibitem{frey2022}
P. Frey, L. Hackl, and S. Rachel, Phys. Rev. B {\bf 106}, L220301 (2022).


\bibitem{hsf6}
S. Moudgalya and O. I. Motrunich, Phys. Rev. X {\bf 12}, 011050 (2022);
D. T. Stephen, O. Hart, and R. M. Nandkishore, arXiv:2209.03966 (unpublished);
D. Hahn, P. A. McClarty, D. J. Luitz, SciPost Phys. {\bf 11}, 074 (2021);
N. Regnault and B. A. Bernevig, arXiv:2210.08019 (unpublished);
D. Vu, K. Huang, X. Li, and S. Das Sarma, Phys. Rev. Lett. {\bf 128}, 146601 (2022).

\bibitem{hsf7}
T. Kohlert, S. Scherg, P. Sala, F. Pollmann, B. H. Madhusudhana, I. Bloch, and
M. Aidelsburger, arXiv:2106.15586 (unpublished).

\bibitem{hsf8}
B. Mukherjee, D. Banerjee, K. Sengupta, and A. Sen, Phys. Rev. B {\bf 104}, 155117 (2021);
P. Brighi, M. Ljubotina, and M. Serbyn, arXiv:2210.5607 (unpublished).

\bibitem{hsf9}
J. Lehmann, P. Sala, F. Pollmann, and T. Rakovszky, arXiv:2208.12260 (unpublished).

\bibitem{hsf10}
A. Chattopadhyay, B. Mukherjee, K. Sengupta, and A. Sen, arXiv:2208.13800 (unpublished).

\bibitem{ghosh2023}
S. Ghosh, I. Paul, and K. Sengupta, Phys. Rev. Lett. {\bf 130}, 120401 (2023).

%\bibitem{SI} See supplementary information for more details.

\bibitem{zhang2022}
S.-B. Zhang, M. M. Denner, T. Bzdu\v{s}ek, M. A. Sentef, and T. Neupert,
Phys. Rev. B {\bf 106}, L121102 (2022).

\bibitem{dias2000}
R. G. Dias, Phys. Rev. B {\bf 62}, 7791 (2000).

\bibitem{rigol1} M. Rigol, V. Dunjko, M. Olshanii, Nature {\bf 452}, 854 (2008).

\bibitem{rigol2}
E. Khatami, G. Pupillo, M. Srednicki, M. Rigol, Phys. Rev. Lett. {\bf 111}, 050403 (2013).

\end{thebibliography}
\end{document}